\documentclass[%
  aps,
  pre,
  superscriptaddress, letterpaper,
  longbibliography,
  onecolumn,
  showpacs,
  showkeys,
  amsfonts, amssymb, amsmath]{revtex4-1}

\usepackage[utf8]{inputenc}
\usepackage[caption=false]{subfig}
\usepackage{hyperref} 

\hypersetup{pdfpagemode={UseOutlines},
bookmarksopen=true,
bookmarksopenlevel=0,
hypertexnames=false,
colorlinks=true, 
citecolor=blue, 
linkcolor=red, 
urlcolor=black, 
pdfstartview={FitV},
unicode,
breaklinks=true,
}
\usepackage{graphicx,amssymb,amsmath}
\usepackage[english]{babel}
\usepackage{grffile}
\usepackage{color}
\usepackage{SIunits}

\usepackage{ulem}
\newcommand{\cor}[1]{#1}

\begin{document}

\title{Generation of weakly nonlinear turbulence of internal gravity waves in the Coriolis facility}
\author{Cl\'ement Savaro, Antoine Campagne, Miguel Calpe Linares, Pierre Augier, Jo\"{e}l Sommeria, Thomas Valran, Samuel Viboud,  Nicolas Mordant}
\email[]{nicolas.mordant@univ-grenoble-alpes.fr}
\affiliation{Laboratoire des Ecoulements G\'eophysiques et Industriels, Universit\'e Grenoble Alpes, CNRS, Grenoble-INP,  F-38000 Grenoble, France}

\begin{abstract}
We investigate experimentally stratified turbulence forced by waves. Stratified turbulence is present in oceans and it is expected to be dominated by nonlinear interaction of internal gravity waves as described by the Garrett \& Munk spectrum. In order to reach turbulent regimes dominated by stratification we use the Coriolis facility in Grenoble (France) which large size enables us to reach regimes with both low Froude number and large Reynolds number. Stratification is obtained by using vertically linearly varying salt concentration and we force large scale waves in a $6\times6\times 1$ m$^3$ domain. We perform time-resolved PIV to probe the space-time structure of the velocity field. We observe a wide band spectrum which is made of waves. Discrete modes are observed due to the square shape of the flow container as well as a continuum part which appears consistent with an axisymmetric superposition of random weakly nonlinear waves. Our observations support the interpretation of turbulence of a strongly stratified fluid as wave turbulence of internal waves although our spectrum is quite different from the Garrett \& Munk spectrum. Weak turbulence proceeds down to a small cutoff length scale (the buoyancy wavelength) at which a transition to more strongly nonlinear turbulence is expected. 
\end{abstract}


\maketitle


\section{Introduction}

Internal gravity waves propagate in the bulk of a fluid which density is non uniform in the vertical coordinate (along the direction of gravity). This situation occurs very generically in geophysical and astrophysical flows \cite{Vallis_2006} in which the fluid is most often subjected to rotation as well. The ocean interior is stratified in density due to variations in temperature and salinity. The atmosphere is stratified due to effects of  temperature. Fluid cores of planets and stars can be similarly stably stratified in some region as well as protoplanetary discs. Internal waves play a major role in the dynamics in particular in the transport of energy and its dissipation through viscosity (kinetic energy) and irreversible mixing (potential energy) but also possibly though coupling with the large scale flow. The contribution of internal waves is of utmost importance for the ocean budget (see for instance \cite{Wunsch_2004,McKinnon_2017}). A major issue to take internal waves into account in ocean models is that they act at small scales and consequently their dynamics cannot be resolved by current global models and their action must be taken into account through parametrization, which is the object of sustained efforts (see for instance \cite{McKinnon_2017}). Serious difficulties arise from the large variety of physical phenomena responsible for the generation and the dissipation of energy carried by the waves but also from the spatial and temporal inhomogeneities of these sources and sinks. Furthermore waves can carry energy over very long distances so that dissipation and mixing can occur at places remote from their generation. 

Among the sources of internal waves in the ocean, one can mention: action of surface waves, wind and storms on the upper surface, turbulence in the upper mixed layer, oscillation of tides on the topography or lee waves radiated by currents on topography  \cite{McKinnon_2017,Polzin_2011}. Dissipation and mixing occurs mostly through overturning of small scale internal waves but also through interaction with the topography~\cite{McKinnon_2017}. Due to the important role played by the bottom topography, dissipation is strongly inhomogeneous in space~\cite{Polzin:1997uy}. Furthermore, action of wind and storms is strongly intermittent in space and time at short term and display also strong seasonal variations~\cite{Polzin_2011}. 

In addition to these sources and sinks, nonlinearity also plays a major role during the propagation of waves in the bulk of the fluid. Indeed nonlinear wave-wave interaction strongly affects the spectrum of the waves by transferring energy to inertial frequencies (low frequencies that are comparable to the local Coriolis parameter) and to small spatial scales at which overturning is more likely to occur. However this assumes a high frequency source of energy whereas near inertial and tidal sources are believed to be dominant~\cite{Polzin_2011}. Again the nonlinear phenomena at play are multiple (see~\cite{Polzin_2011} for a review) and they are based generically on triadic interaction of waves. Such nonlinearities are responsible for the generation of a continuous spectrum of turbulent motions supported by the waves. A well known attempt to describe this spectrum is the Garrett \& Munk (GM) spectrum \cite{Garett_1979} that is based on a synthesis of many oceanographic observations in the 70's so that to build an empirical analytic formula of the wave spectrum. A recent analysis of more recent databases show that the variability of observations is not taken into account by the GM spectrum~\cite{Lvov:2004p1772,Polzin_2011}. Theoretical work suggests rather a family of spectra \cite{Lvov2010} and that the GM spectrum is not a stationary solution of the nonlinear wave equation but may result from a balance between inhomogeneous/nonstationary transport and nonlinear terms~\cite{Polzin_2011}. The theoretical analysis is based on the framework of Weak Turbulence Theory (WTT) which central hypothesis is that the waves are weakly nonlinear~\cite{Nazarenko_2011}. Following this theory, a kinetic equation for the evolution of the internal wave spectrum could be derived~\cite{Lvov:2004p1772,Polzin_2011}. Several theoretical issues have been raised that prevent one to obtain stationary solutions usually referred as Kolmogorov-Zakharov spectra as could be achieved for many other types of waves~\cite{Nazarenko_2011}. These issues are related at the technical level to convergence of interaction integrals and related physically to the locality of nonlinear interactions~\cite{Lvov2010}. Thus the relevance of the Weak Turbulence theory to internal waves remains an open question to a large extent. \cor{Stratified turbulence is an extremely complex setup as the level of nonlinearity can change significantly in scale and the behavior in time and space is somewhat disconnected due to the peculiarity of the internal waves that the frequency of the waves depends only on the propagation direction and not on the wavelength. Furthermore the anisotropy of the flow induces different behaviors in horizontal or vertical directions. Globally one expects a cascade of energy from large to small spatial scale either through wave interactions or stronger nonlinear interactions and a cascade from large to low frequency \cite{Polzin_2011} but a global theory taking into account all aspects of stratified turbulence is still in construction.}

In this article, we focus on the phenomenon of wave-wave nonlinear interaction. The goal is to investigate experimentally if a turbulent regime can be obtained by forcing directly internal waves as in the ocean. Indeed previous experiments~\cite{Augier:2014gv} usually force vorticity rather than waves and use relatively small scale facilities, which impose a small Reynolds number at low Froude number. Our goal is thus to force large scale waves and observe if a cascade of energy to small scale can develop. 

In the laboratory, it is possible to simplify the setup so that to control and simplify the physics of the flow. Experiments are performed in the Coriolis facility (Grenoble, France), which is a large scale facility specifically designed to study stratified flows (with the possible addition of rotation which is not considered here). We directly force waves at large scale and high frequency in a large 3D rectangle domain so that to observe the resulting nonlinear transfers. The main question is wether a state of wave turbulence can be reached. The triadic nonlinear interaction have been observed previously in several experiments but most often at the level of a single triad specifically to highlight the occurence of the Parametric Subharmonic Instability (PSI) in which a single forced wave becomes unstable and gives rise to two daughter waves at lower frequencies close to half the frequency of the mother wave \cite{mcewan_1975,Staquet_2002,Joubaud_2012}. An exception is the experimental work in the group of T. Dauxois in Lyon in a very specific setup involving internal wave attractors \cite{brouzet_2017}. The dispersion relation of internal waves is very specific. For a fluid  with a varying (stable) stratification $\frac{d\rho}{dz}<0$, a plane wave of wave vector $\mathbf k=(k_x,k_y,k_z)$ (with $z$ the altitude, positive to the top) and frequency $\omega$ will follow:
\begin{equation}
\omega^2 = N^2\sin \theta=N^2\frac{k_x^2+k_y^2}{k_x^2+k_y^2+k_z^2}
\label{eq:DR}
\end{equation}
In the above equation $N$ is the Brunt-V\"ais\"al\"a 
\begin{equation}
N= \sqrt{-\frac{g}{\rho_0}\frac{d\bar{\rho}}{dz}}
\label{eq:BV}
\end{equation}
with $\bar{\rho}(z)$ the average density vertical profile, $\rho_0$ is the average density and $\theta$ is the angle between the wave vector $\mathbf k$ and the vertical.
Due to this peculiar dispersion relation, the reflection on an inclined surface does not follow the classical Snell-Descartes law. Indeed the conservation of the frequency imposes the angle to the direction of gravity and the angle to the normal to the surface. A consequence of this feature is that wave beams can be focussed at reflection on an inclined wall (thus by a linear phenomenon) and lead to the concentration of energy on a singular structure called attractor (see \cite{Lam:2009fa,Maas:2011jb}). Dauxois and coworkers take advantage of this energy focussing as for a large enough forcing, the attractor can become unstable (mostly through PSI) and generate nonlinear states of internal waves (see \cite{Joubaud_2012,brouzet_2017,Dauxois:2017ch}). In this configuration energy is injected mostly at very small scales close to dissipative scales which may not be the most efficient way to develop a turbulent cascade. Furthermore their experimental tank is 2D and relatively small. Size is an issue in experiments related to nonlinear internal waves. Indeed, to obtain a flow strongly dominated by gravity effects the Froude number $Fr=\frac{U}{NR}$ needs to be small ($U$ is a velocity scale, $R$ is a length scale). In experiments, the stratification is made with salt and $N$ cannot be increased much beyond typically 1 rad/s. To reach a small $Fr$ then a small velocity and a large size are needed. At the same time, in order to have nonlinear effects, the Reynolds number $Re=\frac{UR}{\nu}$ must be large. Globally this means that $U$ cannot be too small and thus the size must be large to overcome dissipative effects. This is our motivation to use a setup that is significantly larger than previous experiments. Numerical simulations of strongly stratified flows are now possible with ideal periodic boundary conditions (see for instance \cite{Rorai:2015bw,Maffioli:2016bb,Feraco:2018ks,Sujovolsky:2019ix}) but remain challenging in weakly nonlinear conditions due to the large timescale separation between the wave period and the long nonlinear timescale. Most studies do not really analyze specifically the wave dynamics due to a lack of simultaneous space and time resolution in the data processing. Note that flows strongly dominated by rotation support inertial waves which share many similarities (at least at the linear level) with internal gravity waves. Experiments of rotating turbulence have shown to some extent nonlinear regimes of waves \cite{Campagne:2014ef,Campagne:2015fe,Brunet,Yarom:2014df,Yarom:2017ha,Salhov:2019de} but in setups in which waves are not really forced directly but rather through interaction with vortices, which make the analysis more complex.

\section{Experimental setup}

\begin{figure}[!htb]
    \centering
    (a)\includegraphics[width = 8cm]{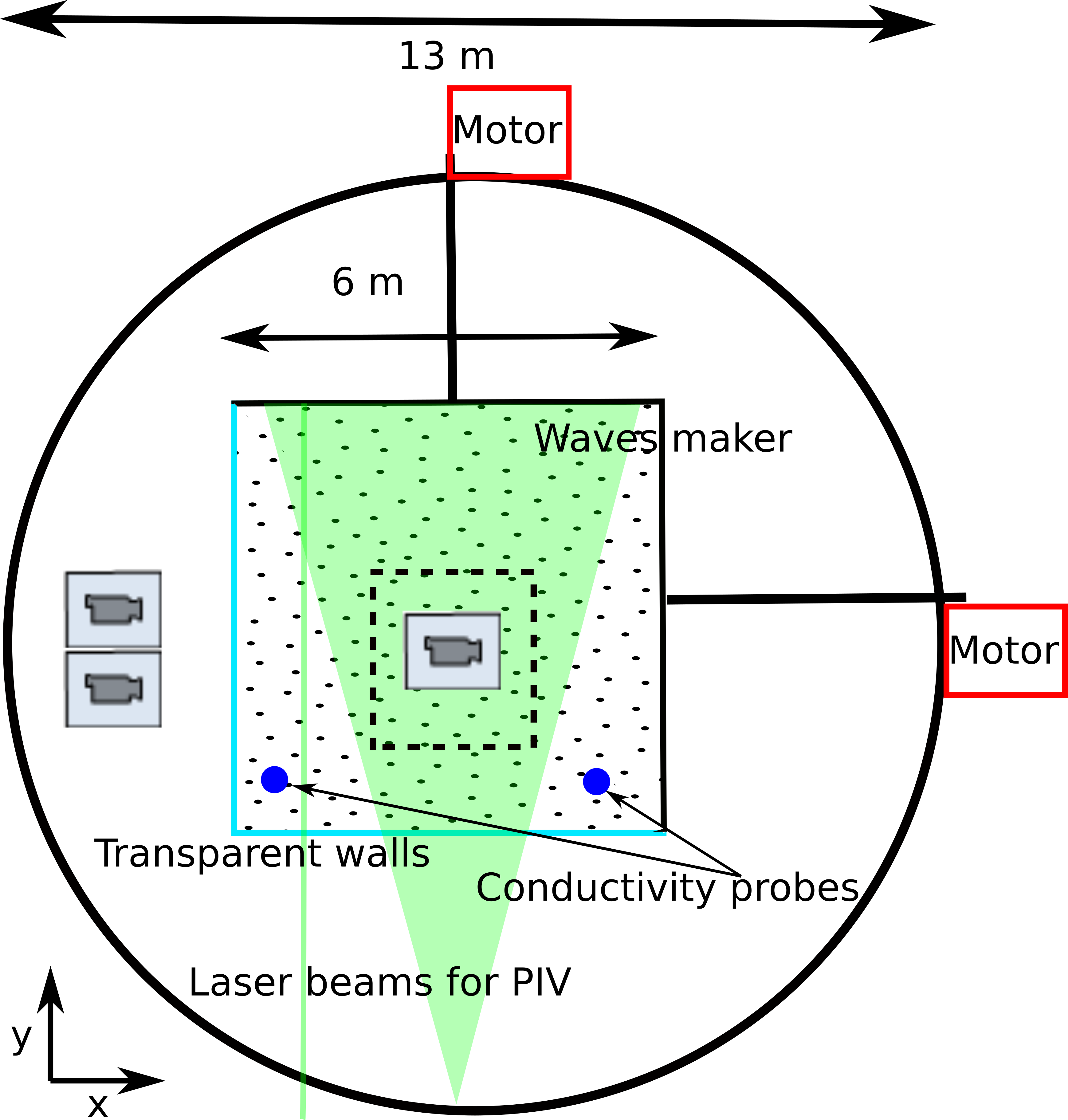} \hspace*{2cm}
    (b)\includegraphics[width = 5cm]{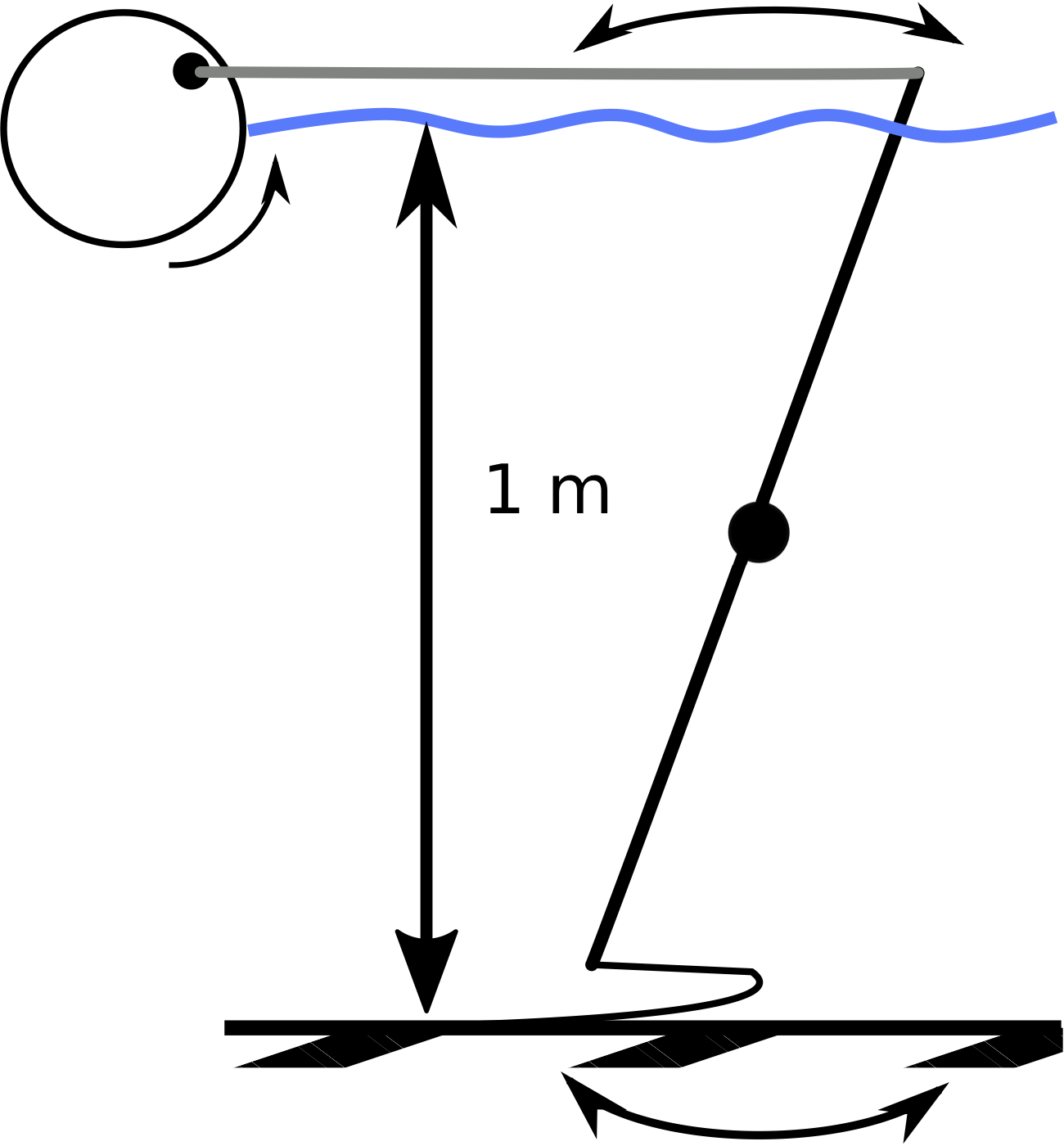}
  \caption{Experimental setup (a) Schematic of experimental setup. The flow is generated in a square domain of side length $L=6$~m inside de 13~m-diameter circular tank of the Coriolis facility with a water depth $H=1$~m.. Two walls are transparent for imaging purposes. The two other walls can oscillate around an axis at mid water depth so that to generate waves. A vertical and a horizontal laser sheet are used for velocity measurement by Particle Image Velocimetry (PIV). See text for more details. (b) Schematic of waves maker. The oscillating panels are driven by a crankshaft system that allows us to tune the amplitude and frequency of oscillation.}
\label{fig:EXP}
\end{figure}

Experiment were carried out in the \unit{13}{\meter} diameter tank of the Coriolis facility in Grenoble (fig.~\ref{fig:EXP}). Inside the tank, we isolate a \unit{6\times6\times1}{\cubic\meter} square domain with two adjacent motionless walls and two oscillating walls used as waves generators. The wave generators oscillate around their mid-height horizontal axis, forcing the mode with one half wavelength vertically (fig.~\ref{fig:EXP}). They are powered by a crankshaft system where the crank is a disk on which a rod can be fixed at different radii to tune the amplitude of forcing oscillations $A$ between 2 and 5~cm (i.e. the top of the panel oscillates horizontally by $\pm A$). The motor rotation rate control the frequency of forcing $\omega_f$. For the data presented in the present article, the frequency is randomly varied in a narrow band of width $5 \%$ centered on $\omega_f/N=0.7$.  Plane waves oscillating at this frequency would propagate with an angle $\theta\approx \pi/4$. One can build a forcing Reynolds number by using $H$ as the length scale (which is half the vertical wavelength forced by the wavemaker) and the maximum horizontal velocity $0.7NA$ of the top of the oscillating panel (assuming a perfect efficiency of the wavemaker) so that $Re_f=0.7NHA/\nu$ which lies between $8\,10^3$ and $2\,10^4$ with the parameters of the experiment (see table~\ref{tab}). 

\begin{figure}[!htb]
 	\centering
	\includegraphics[width=10 cm]{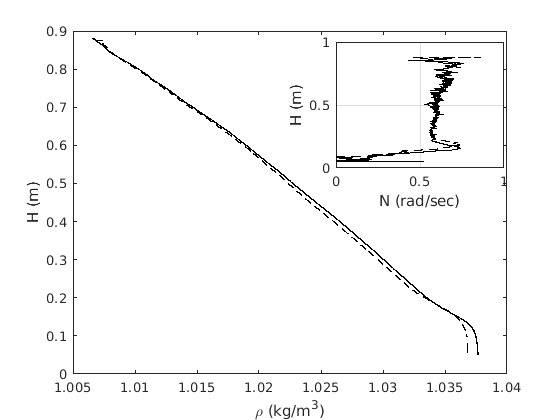}
	\caption{Example of density profiles before (line) and after (dashed line) experiment C ($A=4$~cm). The duration of the experiment is 5 hours and the profile was measured after 2 extra hours of velocity decay. Insert: local Brunt-V\"ais\"al\"a frequency $N(z)$ for both cases.}\label{fig:profile}
 \end{figure}

The various walls are sealed with canvas so that the flow domain remains isolated from water outside the domain. The whole tank is filled up by a \unit{1}{\meter} deep stable linear stratification of salt water so that the Brunt-V\"ais\"al\"a frequency is uniform and is close to $N=0.6$~rad/s (which corresponds to a minimum wave period of 10~s). Four pre-calibrated conductivity probes monitor the density field. Two probes are fixed at different depths (20 and 80 \centi\meter) and two are mounted on a vertical profiler to record profiles of stratification before and during experiments. \cor{The turbulent flow induces some mixing. For a linear profile, if the mixing is statistically homogeneous in space, the linear profile is unchanged and the mixing is only visible through the existence of homogeneous layers at the very top or very bottom of the tank. These mixed layers when getting thicker reduce the effective depth of the stratification. The development bottom layer is visible in fig.~\ref{fig:profile}. The top layer is usually thinner and not visible here due to technical difficulties to measure near the surface.} The mixing process remains slow enough to be neglected for the duration of one experiment. However as several experiments are performed one after another, the effective depth $H$ varies slightly and could be measured from the density profile if needed. To perform Particle Image Velocimetry (PIV) measurement, the fluid is seeded with particles whose density spans that of the fluid stratification. In this way particles are scattered all over the fluid volume and do not settle down under gravity. The particles were obtained by heating slightly polystyrene beads. Thin layers are left free of particles at the top and the bottom of the fluid to avoid sedimentation at the bottom and floating particles at the top due to the presence of the mixed layers. Two distinct PIV setups are installed in the experiment :  bidimensional - two components (2D-2C) in a horizontal and a vertical plane:
\begin{itemize}
\item a horizontal light sheet is generated using a \unit{25}{\watt}, \unit{532}{\nano\meter} CW laser passing through a Powell lens. A 12 Mpixels camera is fixed 4 meters above water level. Its field of view is \unit{2.5\times2.1}{\meter\squared} (at mid-height without water). The camera records images series at constant frame rate (3.3~frames/s), from which time-resolved velocity fields are obtained by  2D PIV. A scanning mirror can be used to vertically move the laser sheet so that to obtain a 3D-2C measurement.
\item a vertical light sheet is generated with a fast oscillating mirror using a \unit{5}{\watt}, \unit{532}{\nano\meter} CW laser beam. Two side by side cameras pointing downside to a \unit{45}{\degree} inclined underwater mirror  record the whole water column (through a transparent sidewall) and a total of \unit{2}{\meter} in the $y$ direction with some overlap. We obtain thus the components $v$ and $w$ in a plane of constant $x$, about \unit{1.5}{\meter} from the transparent wall. A difficulty associated with this PIV plane is that it is more sensitive to fluctuations of the index of refraction than the previous configuration due to the much longer propagation distance of light in water between the laser sheet and the camera. At the strongest forcing, it results in a blurring of the images due to localized strong mixing events on the path of light. 
\end{itemize}

\begin{table}    
\begin{tabular}{p{1.5cm}p{1cm}p{1.2cm}p{1cm}p{1cm}p{1cm}p{1cm}p{1cm}p{1cm}p{1cm}p{1cm}p{1cm}p{1cm}}
\hline\noalign{\smallskip}
Dataset & $A$ & duration & fields& $u_f$ & $Re_f$ & $Fr_f$ & $Re_{bf}$ & $\sigma_u$ & $\sigma_v$ & $\sigma_w$  & $k_b$\\
 & [cm] & [min] &  & [cm/s] &&& & [cm/s] & [cm/s] & [cm/s] &  [m$^{-1}$]\\
\noalign{\smallskip}\hline\noalign{\smallskip}
A & 2 & 240 & v \& h & 0.8 & 8400 & 0.014 & 1.6 & 0.4 & 0.6 & 0.8 &  70\\
B & 3 & 120 & v \& h & 1.3 & 13000 & 0.021 & 5.6 & 0.7 & 0.8 & 1.0 &  50\\   
C & 4 & 120 & v \& h & 1.7 & 17000 & 0.028 & 13 & 1.0 & 1.4 & 1.1 &  35\\       
D & 5 & 30 & h & 2.1 & 21000 & 0.035 & 26 &1.5 & 1.4 & n/a & 28 \\       
\noalign{\smallskip}\hline\noalign{\smallskip}
\end{tabular}
\caption{Parameters of the experiments. $A$ is the amplitude of oscillation of the wavemakers. The `fields' column specifies which of the PIV fields are available (`h' stand for the horizontal measurement and `v' for the vertical one). $u_f$ is the velocity of the top of the panels of the wavemakers providing a typical horizontal velocity $u_f=0.7NA$ (the frequency of oscillation is close to $0.7N$). The forcing Reynolds number is $Re_f=u_fH/\nu$ with $\nu$ the viscosity of water. The forcing Froude number is defined as $Fr_f=u_f/NH=0.7A/H$. The forcing buoyancy Reynolds number is $Re_{bf}=Re_fFr_f^2=0.34A^3N/H\nu$. $\sigma_u$, $\sigma_v$ and $\sigma_w$ are the {\it rms} values of the velocity along the $x$, $y$ and $z$ axes respectively. $k_b$ is the buoyancy wave number estimated as $k_b=N/u_f$.}
\label{tab}   
\end{table}

\begin{figure}[!htb]
	\centering
	\includegraphics[width=12cm]{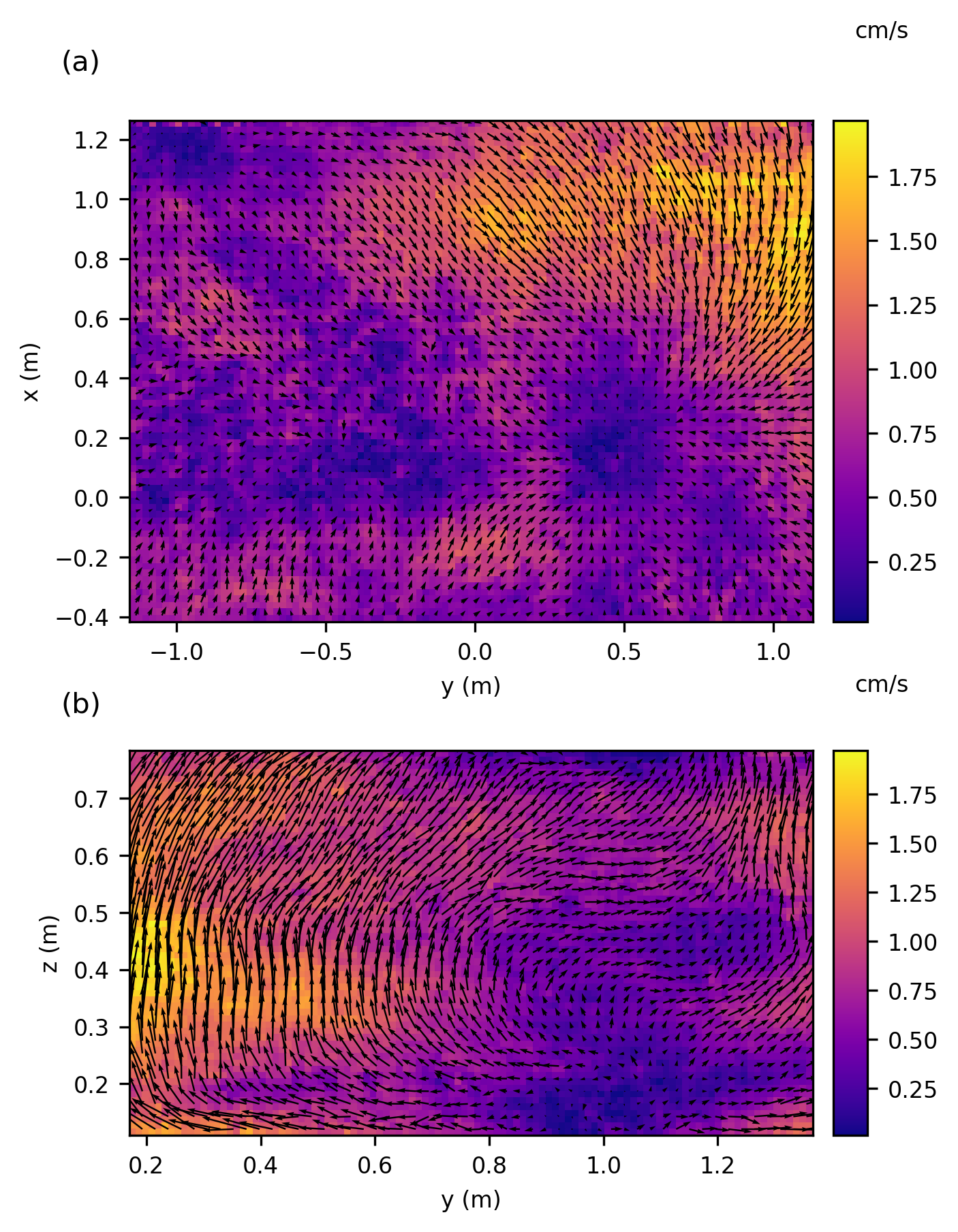}
	\caption{\cor{Example of instantaneous velocity field in (a) horizontal and (b) vertical  planes for experiment A. The color indicates the magnitude of the two component velocity measurements in each case.}}
	\label{fig:Quiver}
\end{figure}

The sets of images are processed by the locally developed FluidImage PIV software \cite{Mohanan:2019ko} to calibrate the images and extract the velocity field. The spatial resolution of the PIV is 1~cm for the vertical laser sheet and 2~cm for the horizontal PIV. Parameters of the various experiments are given in table~\ref{tab}. \cor{It can be seen in particular that the Froude number is very small, of order $10^{-2}$ which means that the flow at large scale is weakly nonlinear. Examples of snapshots of PIV fields are shown in fig.~\ref{fig:Quiver}.}

\section{Frequency analysis}
\label{F_ana}

\begin{figure}[!htb]
 	\centering
	\includegraphics[width=12cm]{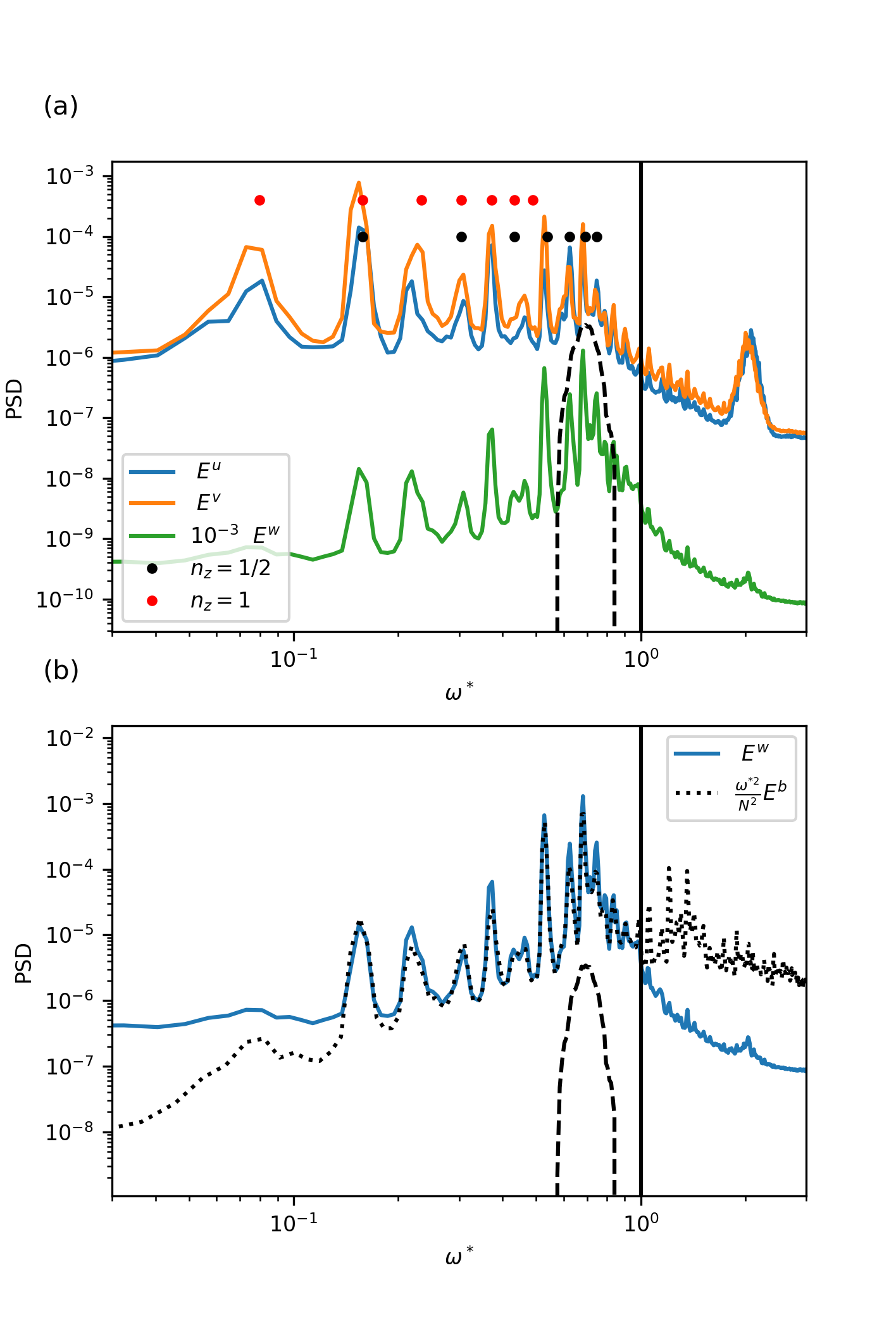}
	\caption{(a) Power spectral density (PSD) of all components of velocity, with $A = \unit{2}{\centi\meter}$. The dots corresponds to the frequency of 2D modes (see eq. (\ref{modes}) and text). $\omega^\star$ is the normalized frequency $\omega^\star=\omega/N$ (b) PSD of the vertical velocity compared to that of the local density (see text for details on the normalization factor). The spectra have been computing using the Welch method with a temporal window duration of \unit{1229}{\second} and using a Hanning window. The spectra are also averaged over all positions available in the PIV field for $w$ and over the two fixed conductivity probes for $\rho$. The dashed line shows the distribution of instantaneous rotation frequencies sent to the motors that drive the oscillating panels.}\label{fig:PSDUVW}
 \end{figure}

Examples of frequency power spectra density of the velocity fields and density are shown in fig.~\ref{fig:PSDUVW} for the smallest forcing amplitude $A=2$~cm. Here $u$ and $v$ are obtained with horizontal PIV and $w$ with vertical PIV. Frequency is normalized with the Brunt-V\"ais\"al\"a frequency : $\omega^* = \omega/N$ (thus waves can exist only for $\omega^*\leq 1$). The frequency spectra are computed using the standard Welch method in which a discrete Fourier transform is performed on successive segments of data of finite duration with a 50\% overlap. Furthermore the velocity spectra are averaged over all space points available. The spectra of $u$ and $v$ are very similar as expected from the square symmetry of the setup. The main feature of the spectra is a rather flat continuum for $\omega^*<0.8$ with several sharp peaks. The spectrum of $w$ show similar peaks but the continuum part is slightly increasing with $\omega$. For $\omega^*>1$, the spectra are fast decaying before becoming flat at $\omega^*>3$ when reaching the noise level. The fast decay for $\omega^*>1$ is consistent with the fact that the frequency of gravity waves cannot be greater than $N$ as can be concluded from the dispersion relation (\ref{eq:DR}). This observation is thus a first hint that much of the energy could be due to waves. Furthermore the peaks can be associated with mode frequencies due to the finite size of the tank. For internal waves in a parallelepipedic  box, the discrete modes have a very standard spatial structure, \cor{with, for example, the $u$ component of velocity being: 
\begin{equation}
u(x,y,z)\propto \sin(2\pi n_x x/L)\cos(2\pi n_y y/L)\cos(2\pi n_z z/H)\, ,
\end{equation}}
where $n_x$, $n_y$ and $n_z$ are multiples of $1/2$ (an integer number of half wavelengths must fit in the box) . The possible discrete frequencies are then
\begin{equation}
\omega^*=\dfrac{1}{\sqrt{1+\dfrac{(n_z/H)^2}{(n_x^2+n_y^2)/L^2}}}\, .
\label{modes}
\end{equation}
Fig.~\ref{fig:PSDUVW}(a) shows the position of the frequencies for $n_z=1/2$ and $1$ (i.e. modes with $1/2$ and $1$ wavelength in the height $H$ of the tank), $n_y=0$ and $n_x=1/2\times(1\cdots 8)$. This corresponds to the frequencies of 2D modes bouncing back and forth between parallel walls. It can be seen that all the observed peaks correspond indeed to these modes. The 3D modes with both $n_x$ and $n_y$ being non zero are not observed as strong peaks, possibly because they have a more complex spatial structure that is damped more by viscosity. \cor{Note that each frequency is infinitely degenerated in (\ref{modes}) (for instance by multiplying all integers $n_i$ by the same integer) and furthermore it is possible that modes with distinct geometries are degenerated as well, such as 3D modes having the same frequency than 2D modes (this is indeed observed as shown in next section).} The lower frequency modes are excited through nonlinearity from the forcing in both $x$ and $y$ direction. Indeed the modes directly excited by the forcing at $\omega^\star\sim0.7$ have similar horizontal and vertical wavelengths i.e. $n_z=n_x/6=1/2$. There is then a transfer of energy to small frequencies as well as a transfer to smaller vertical scales as some modes correspond to smaller vertical wavelengths that the mode $n_z=1/2$ which is directly forced by the wavemaker. The peaks of the spectrum are thus associated with waves which are likely to be weakly nonlinear as the frequency remains close to that of the linear waves. Furthermore the triplets of waves modes that are the closest to be in triadic resonant condition (table (\ref{tab:triad})) appear to be the most prominent peaks in the spectrum.

\begin{table}    
\begin{tabular}{|p{1cm}|p{2.5cm}|p{2.5cm}|p{2.5cm}|p{2.5cm}|p{2.5cm}|p{2.5cm}|}
\hline
$\delta \omega$&0.0027 & 0.0067& 0.0106&0.0109&0.0115&0.0122\\
\hline
$\omega^*$ & (0.16,0.53,0.69)&(0.08,0.53,0.62)&  (0.16,0.37,0.53) &(0.08,0.62,0.69)&(0.30,0.43,0.74)&(0.37,0.43,0.78)\\
$2n_x$ &  (2,4,6) &(1,4,5) & (1,5,4)&(1,5,6)&(4,3,7)&(5,3,8)\\   
$n_z$ &   (1,1/2,1/2)&(1,1/2,1/2)&(1/2,1,1/2)&(1,1/2,1/2)&(1,1/2,1/2)&(1,1/2,1/2)\\          
\hline
\end{tabular}
\caption{Triplets of mode defined by eq. (\ref{modes})) that are the closest to triadic resonant condition, ie $\sum \pm \omega^* = \delta \omega$ and $\sum \pm n_y=\sum \pm n_z = 0$}
\label{tab:triad}   
\end{table}

In the framework of the Boussinesq approximation, the equations of motion can be written~\cite{Vallis_2006}:
\begin{eqnarray}
\frac{\partial \mathbf u}{\partial t}+(\mathbf u\cdot\nabla)\mathbf u&=&b\mathbf e_z-\frac{1}{\rho_0}\nabla p+\nu\Delta \mathbf u\\
\frac{\partial b}{\partial t}+(\mathbf u\cdot\nabla)b&=&-N^2w +\kappa \Delta b
\end{eqnarray}
where $b=\frac{\delta \rho}{\rho_0}g$, with $\delta \rho$ being the density departure from the linear density stratification and $p$ the pression variation from the hydrostatic profile.
The linearized scalar advection equation can be written (neglecting molecular diffusion) 
\begin{equation}
\frac{\partial b}{\partial t}=-N^2 w\, .
\end{equation}
Thus, for weakly non linear regimes, one expects the spectrum of $b$ to follow 
\begin{equation}
E^w(\omega)=E^b(\omega)\omega^2/N^4\, . 
\end{equation}
Fig.~\ref{fig:PSDUVW}(b) shows the comparison between the lhs and rhs terms. One sees that indeed for $\omega<N$ the two curves are very close. For $\omega>N$, the curves are well separated. This observation is consistent with a flow dominated by weakly non linear waves that can exist only for $\omega\leq N$. Note that one reason for which the agreement is not perfect could be that the measurements of velocity and density are not done at the same place in the experiment. The velocity is measured on a large area at the center of the domain while the density is measured in 2 points closer to the wall.

\begin{figure}[!htb]
 	\centerline{\includegraphics[width=12 cm]{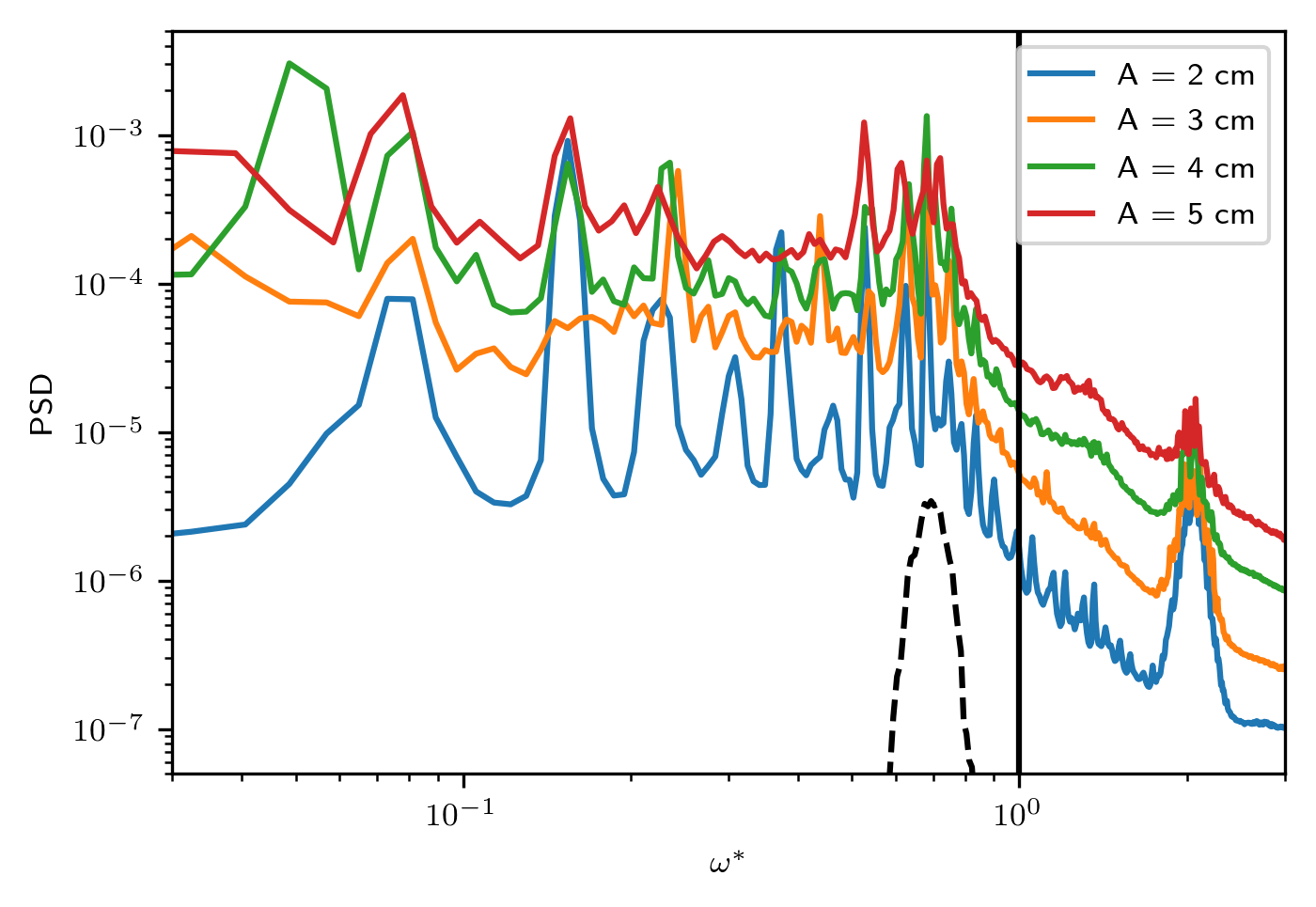}}
 	\caption{Power spectra density of the horizontal velocity (sum of the PSDs of $u$ and $v$) as a function of $A$.}\label{fig:psdamp}
 \end{figure}
Fig.~\ref{fig:psdamp} shows the evolution of the spectrum of the horizontal velocity when the amplitude of oscillation of the wavemakers is increased from 2 to 5~cm. The strongly peaked spectrum observed at the weakest forcing evolves to a more continuous spectrum in which only a few peaks remain visible. The continuum part of the spectrum increases with the forcing while the discrete part of the spectrum remains globally unchanged (except for the first peak at $\omega^*\approx 0.08$). There is an evolution from a discrete sort of turbulence towards a much more continuous spectrum. This is generically expected in the framework of weak turbulence when the nonlinear effects overcome the finite size effets \cite{Lvov:2006p1046,Kartashova:1994p1074}. This phenomenon has been observed also for instance in experiments of weak turbulence of a vibrating plate~\cite{Mordant:2010p1038}. 

A test to check if the continuum is made of waves is to take advantage of geometric properties of internal waves. Let us consider a given linear internal wave of wavevector $\mathbf k=k(\sin\theta\cos\phi,\sin\theta\sin\phi,\cos\theta)$, in spherical coordinates, with $\phi$ the azimuthal angle. The velocity components oscillate in a vertical plane containing $\mathbf k$ and the velocity is orthogonal to $\mathbf k$~\cite{Staquet_2002}. Thus one has $v(\mathbf k,\omega)=a(\mathbf k,\omega)\cos\theta\sin\phi$ and $w(\mathbf k,\omega)=\pm a(\mathbf k,\omega)\sin \theta$ with $a(\mathbf k,\omega)$ the amplitude of the given wave. Let us now assume that the total motion is made of a random superposition of statistically independent linear plane waves, which are axisymmetric around the vertical axis and which follow the linear dispersion relation. Then the statistics of $a$ depends only on $k=|\mathbf k|$ and $\omega$ (that gives the dependency in $\theta$ through the dispersion relation). One can first average over realizations so that
\begin{equation}
\langle |w(\mathbf k,\omega)|^2\rangle=\langle |a(\mathbf k,\omega)|^2 \rangle \sin^2 \theta
\end{equation}
and 
\begin{equation}
\langle |v(\mathbf k,\omega)|^2\rangle=\langle |a(\mathbf k,\omega)|^2 \rangle \cos^2\theta\sin^2\phi
\end{equation}
with $\langle |a(\mathbf k,\omega)|^2\rangle \cor{= A(k,\omega)}$ depending only on $k$ and $\omega$.

We can now sum over the angle $\phi$ so that
\begin{equation}
\int_0^{2\pi} \langle |w(\mathbf k,\omega)|^2\rangle d\phi =2\pi \cor{A(k,\omega)} \sin^2 \theta
\end{equation}
and 
\begin{equation}
\int_0^{2\pi} \langle |v(\mathbf k,\omega)|^2\rangle d\phi=\pi\cor{A(k,\omega)}  \cos^2\theta
\end{equation}
By further summing over $k$ one obtains the frequency spectrum $E^w(\omega)$ and $E^v(\omega)$ so that the ratio of the two gives
\begin{equation}
\frac{E^w(\omega)}{E^v(\omega)}=\frac{2\sin^2\theta}{\cos^2\theta}=\frac{2(\omega^*)^2}{1-(\omega^*)^2}\, ,
\label{eq:ratio}
\end{equation}
which is due to geometric constraints imposed by the structure of the waves at a given frequency.

\begin{figure}[!htb]
	 	\centerline{\includegraphics[width=12 cm]{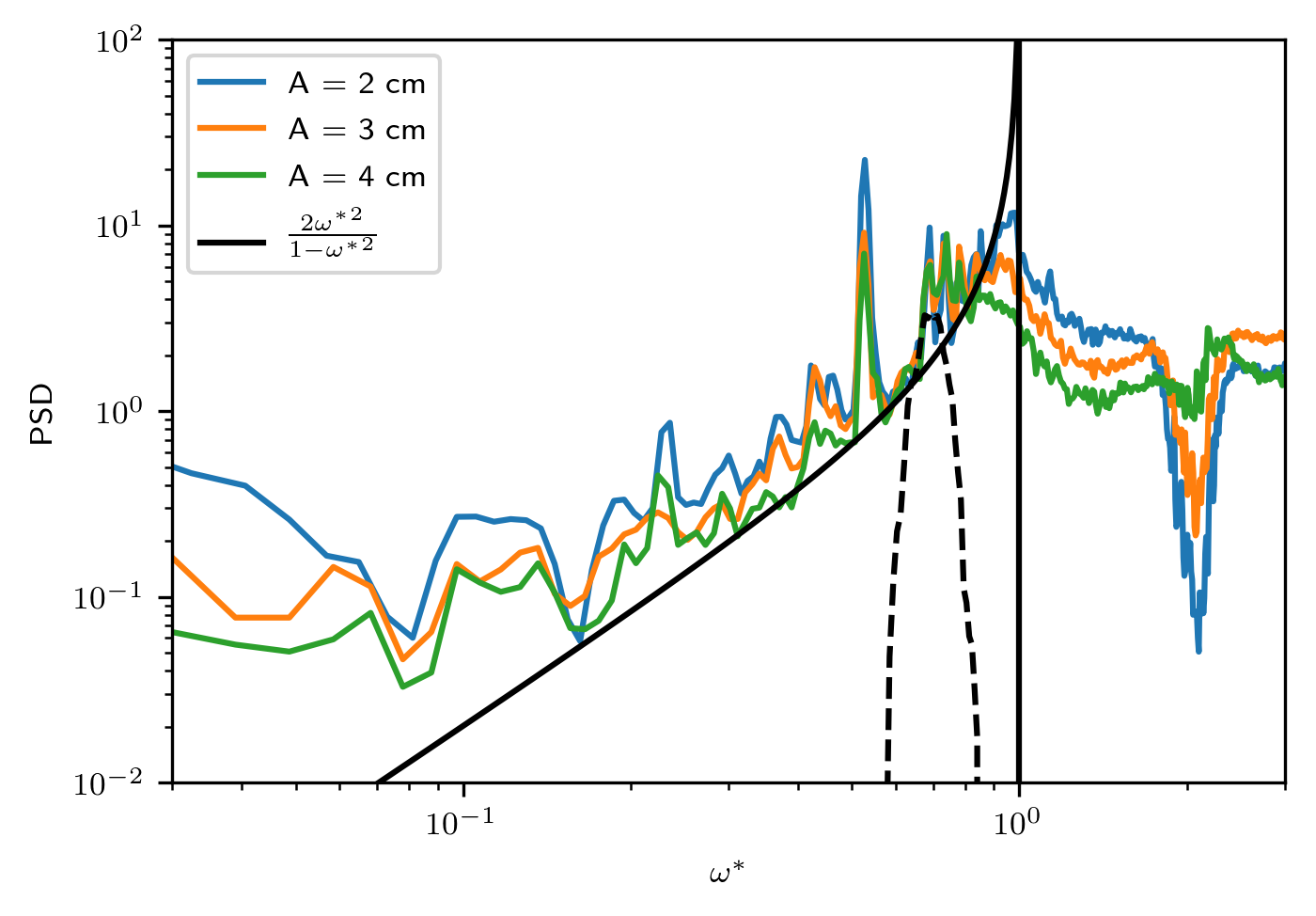}}
 	\caption{Ratio between power spectra density of $w$ and $v$ for various values of $A$. The continuous line shows the rhs of (\ref{eq:ratio}). The dashed line is the forcing frequency range.}\label{fig:ratio}
 \end{figure}
Fig.~\ref{fig:ratio} shows the ratio $\frac{E^w(\omega)}{E^v(\omega)}$ of the measured spectra for the same dataset as the previous figure as well as higher forcing intensities, together with the rhs of (\ref{eq:ratio}) (black line). One sees that indeed the ratio of the spectra are very close to the prediction of the simple axisymmetric model for both the continuum and the peaks (except for one peak at $\omega^*$ close to 0.5). It again suggests strongly that our flow is indeed an axisymmetric superposition of weakly nonlinear waves (except for the 2D modes at the peaks of the frequency spectrum). The experimental lines are actually getting slightly closer to the model prediction when the forcing is increased.

\section{Spatiotemporal analysis}

\begin{figure}[!htb]
	\centering
	\includegraphics[width=12cm]{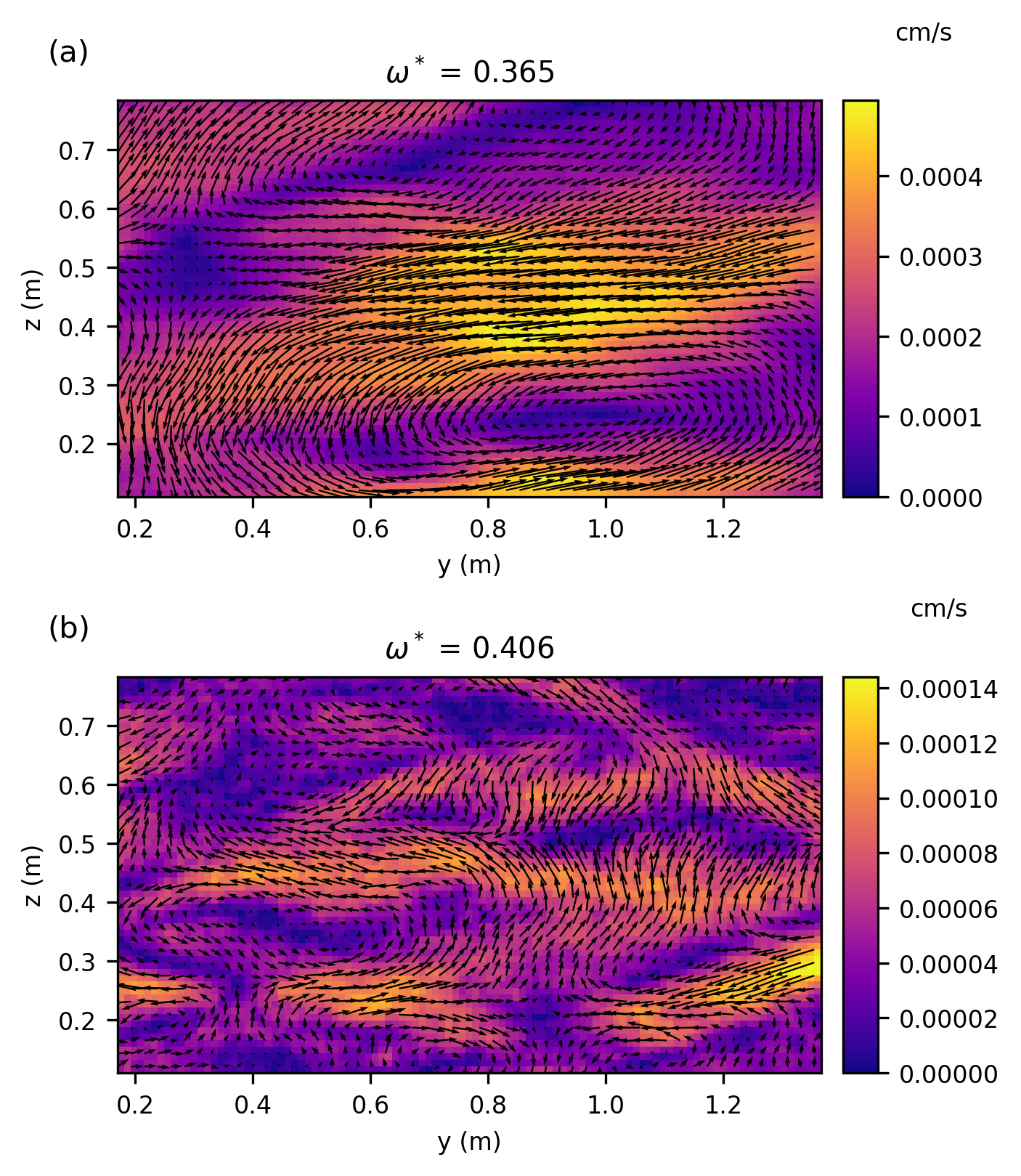}
	\caption{\cor{Filtered velocity field (real part of the Fourier transform) in vertical plane (a) on a frequency peak and (b) at frequency chosen between peaks. Frequencies are specified in the title.}}
	\label{fig:Quiver_filtered}
\end{figure}

\cor{Fig.~\ref{fig:Quiver_filtered} shows the filtered velocity field both on a frequency peak and in the continuum between peaks. The flow structure is clearly different between these two cases, with much larger scale on the frequency peak.}
In order to probe more deeply the wave structure, it is necessary to study the statistics of the fields both in space and time. 

A first analysis is inspired from the work by Campagne {\it et al.} \cite{Campagne:2015fe} in which they performed a Fourier analysis in time but a correlation analysis in space. 
Here we define the correlation of $u$ as:
\begin{equation}
C^u(\mathbf r,\omega)=\frac{\langle u(\mathbf R_0+\mathbf r,\omega)u^\star(\mathbf R_0,\omega)+c.c.\rangle}{2\langle |u(\mathbf R_0,\omega)|^2\rangle}\, ,
\label{eq:2ptcorelation_h}
\end{equation}
where $\mathbf r$ lies in the horizontal or vertical plane for respectively horizontal and vertical correlation. The average is performed as a Welch method with average in time over successive temporal windows of duration $T=1229$~s (with 50\% overlap and a Hanning window) as well as an average over $\mathbf R_0$. $c.c.$ stands for ``complex conjugate'' and $\cdot^\star$ stands for the complex conjugaison operation. 

\begin{figure}[!htb]
 	\centering
	\includegraphics[width=16 cm]{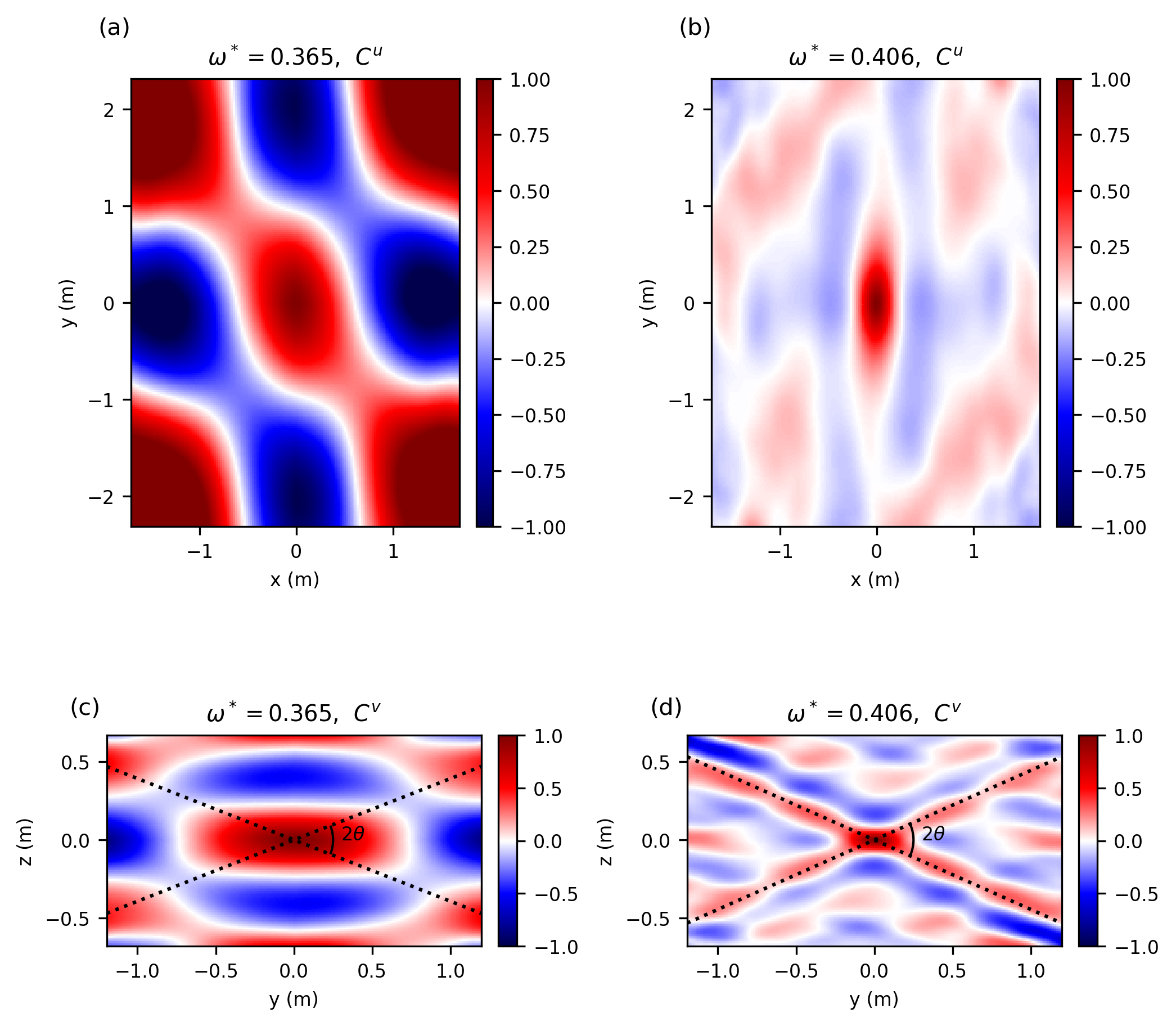} 
	
	\caption{Images of space-frequency correlations for $A=2$~cm. (a) \& (c) are taken at the frequency $\omega^*=0.365$ for which there is a strong peak in the spectrum. (b) \& (d) correspond to $\omega^*=0.406$ in the continuum. (a)\&(b): horizontal plane correlations for one component of velocity $C^{u} (x,y,\omega)$. (c)\& (d): vertical plane correlations for horizontal component of velocity $C^{v} (y,z,\omega)$. $\theta$ is the angle obtained from the dispersion relation at the given frequency.}
	\label{fig:rom}
\end{figure}

\begin{figure}[!htb]
 	\centering
	\includegraphics[width=16cm]{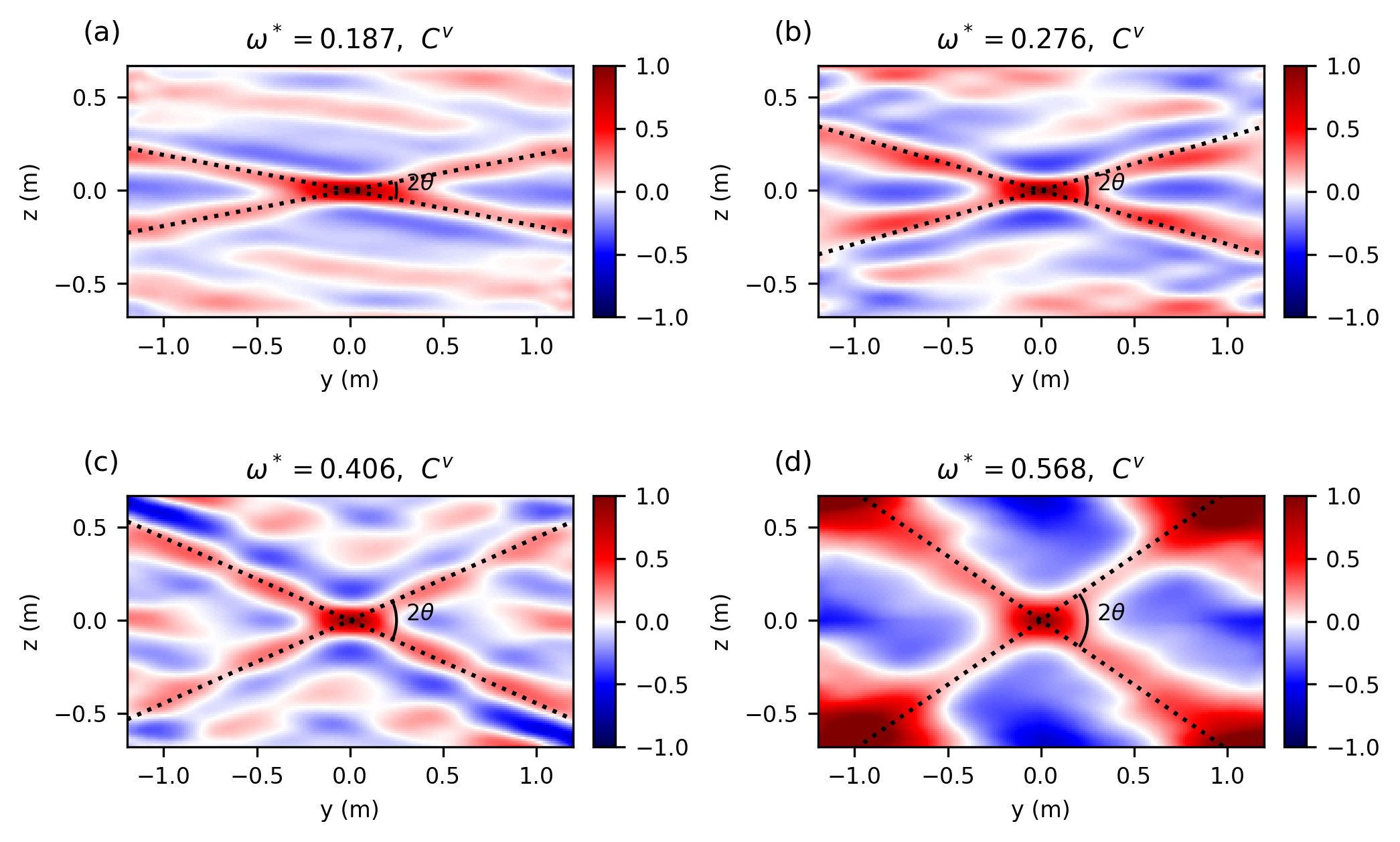} 
 	\caption{Images of space-frequency vertical correlations for horizontal component of velocity $C^{v} (y,z,\omega)$ for various frequencies chosen between peaks. The value of the frequency is given in the title.}
	\label{fig:rom1}
\end{figure}

Fig.~\ref{fig:rom} shows pictures of the correlations of one horizontal component of velocity in both horizontal ((a)\&(b)) and vertical ((c)\&(d)) planes and for a frequency chosen either on a peak of the spectrum ((a)\&(c)) or in between peaks ((b)\&(d)), i.e. in the continuum. The correlations observed for the two frequencies are very different. Concerning this specific value of the peak frequency, we expect to observe a superposition of 2D modes defined as: $(n_x,n_y,n_z) = (0,5/2,1)$ and $(n_x,n_y,n_z) = (5/2,0,1)$ (fig.~\ref{fig:PSDUVW}), so with a horizontal wavelength $\lambda =L/(5/2)= 2.4\,\meter$ and vertical wavelength equal to the water depth $H$. \cor{However the horizontal correlation in fig.~\ref{fig:rom}(a) is instead dominated by the 3D mode $(n_x,n_y,n_z) = (4/2,3/2,1)$, with horizontal wavelength $\lambda_x = 3\meter$ and $\lambda_y = 4\meter$. This 3D mode is degenerated with the 2D modes at the considered frequency. It remains a very large scale mode. Even though the frequency corresponds to that of a 2D mode for some reason energy is preferably transferred to the 3D mode. Note also that the same 3D mode obtained by rotation of $\pi/2$ is not visible. The reason why a single mode is dominating is most likely due to details of the geometry of the experimental setup and on the process of nonlinear saturation. For other peaks, the 2D modes are indeed observed.} 
The vertical correlation at the same frequency (fig.~\ref{fig:rom}(b)) shows similar features, the dominant vertical wavelength is close to 1~m which is equal to the depth $H$ of the domain. For the frequency chosen in the continuum, the horizontal correlation looks very different (fig.~\ref{fig:rom}(b)): the correlation displays a narrow peak at the origin with a fast decay to values close to zero (within the statistical convergence of the estimator). The peak is wider in the transverse ($y$) direction: the mid-height half-width of the peak is $\unit{0.38}{\meter}$ in transverse direction and $\unit{0.14}{\meter}$ in the longitudinal direction. The vertical correlation (fig.~\ref{fig:rom}(d)) has a very peculiar St.~Andrew cross shape with a peak in the middle. The vertical half height width of the central peak is even smaller ($\unit{0.05}{\meter}$ in the $z$ direction and $\unit{0.14}{\meter}$ in the $y$ direction). A positive (red) and negative (blue) cross is visible away from the peak. The dotted lines have an angle $2\theta$ between each other where $\theta$ is the angle corresponding to the chosen frequency through the dispersion relation. This feature is not specific of the chosen frequency: Fig.~\ref{fig:rom1} shows the same estimator for three other values of the frequency. In all cases a similar cross pattern can be observed.

\begin{figure}[!htb]
 	\centering
	\includegraphics[width=16cm]{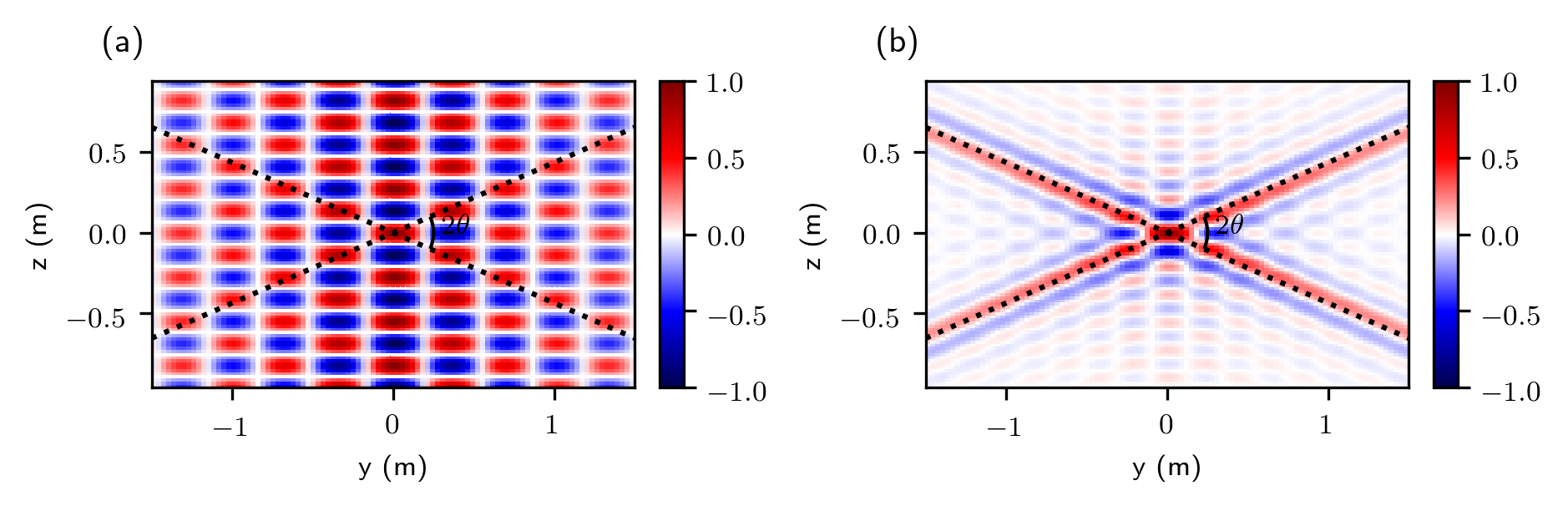}
 	\caption{Images of space-frequency correlations $C^{v}(y,z,\omega)$ for an isotropic superposition of independent linear waves for $\omega^\star=0.41$. (a) for a single wavenumber $k=\unit{25}{\rad.\reciprocal\meter}$ (b) for a uniform superposition of $k$ between $5$ and $\unit{40}{\rad.\reciprocal\meter}$. See text for details.}\label{fig:model}
\end{figure}

In order to interpret the structure of the correlations we use the same model of random, axisymmetric superposition of independent linear plane waves used above in the spirit of Campagne {\it et al.} \cite{Campagne:2015fe}. At a given frequency, the velocity field can be written as
\begin{equation}
\mathbf u(\mathbf r,\omega)=\int \mathbf a(\mathbf k,\omega) e^{i\mathbf k\cdot \mathbf r} d\mathbf k
\end{equation}

The two-point correlation of the velocity is thus
\cor{\begin{equation}
\langle \mathbf u(\mathbf R_0+\mathbf r,\omega)\mathbf u^\star(\mathbf r,\omega)+c.c.\rangle=\iint \langle\mathbf a(\mathbf k_1,\omega)\mathbf a^\star(\mathbf k_2,\omega)\rangle e^{i(\mathbf k_1-\mathbf k_2)\cdot \mathbf R_0+i\mathbf k_1\cdot \mathbf r} d\mathbf k_1d\mathbf k_2 +c.c.
\end{equation}}
The velocity field is homogeneous in space so that the correlation does not depend on $\mathbf R_0$. An average over $\mathbf R_0$ can be performed that provides a $\delta(\mathbf k_1-\mathbf k_2)$ in the integral so that the correlation, normalized like eq. (\ref{eq:2ptcorelation_h}) can be written:
\begin{equation}
C(\mathbf r,\omega)=\frac{\int \langle|\mathbf a(\mathbf k,\omega)|^2\rangle \cos(\mathbf k\cdot \mathbf r) d\mathbf k}{\int \langle|\mathbf a(\mathbf k,\omega)|^2\rangle d\mathbf k}
\end{equation}
Again, because of axisymmetry, $\langle|\mathbf a(\mathbf k,\omega)|^2\rangle$ does not depend on the azimuthal angle $\phi$ and only on $k$ and $\theta$. $\theta$ is actually imposed by the frequency through the dispersion relation. 
For a single wavenumber $k$, Campagne et al. \cite{Campagne:2015fe} showed that the correlation can be rewritten as (after adapting the dispersion relation from inertial to internal waves)
\begin{equation}
C(\mathbf r,\omega)= \cos(kz\sqrt{1-{\omega^\star}^2})J_0(kr_\bot\omega^*)
\end{equation}
where $J_0$ is the Bessel function of first kind, and $r_\bot$ is the length of the projection of $\mathbf r$ in the horizontal plane. In this expression $C$ is the full velocity field correlation. In our case we have only two components of the velocity vector projected in a plane. With the same assumptions one can compute the expression of the correlations of individual components projected on different planes (see appendix).

Figure \ref{fig:model}(a) shows the computation of this model for horizontal velocity in vertical plane (eq (\ref{eq:A:CUvertlong})) for a single scale $k= 25$~rad$.\meter^{-1}$. The observed network is consistent with what is obtained in fig.~\ref{fig:rom}(b) (although at a larger scale) at a frequency corresponding to a peak in the frequency spectrum. By contrast, fig.~\ref{fig:model}(b) shows the case of a broadband superposition of scales in the interval $5<k<40~$~rad$.\meter^{-1}$ assuming arbitrarily that all waves have the same amplitude (the choice of the spectrum of the amplitude of the waves does not change qualitatively the picture, see appendix). The observed St.~Andrew cross is very similar to that shown in fig.~\ref{fig:rom}(d). It strongly suggests that the continuum part of the frequency spectrum corresponds to a random ensemble of propagating waves with a rather broad range of wavelengths. These waves corresponds both to a transfer of energy to lower frequencies than the forcing and to smaller length scales as the forcing is expected to be efficient at wave lengths close to 2 meters. Here the width of the St Andrew cross for $A=2$~cm shows that wavelengths about 10 times smaller are present in the flow. The width of the cross is related to the high wavenumber cutoff of the model (see appendix).

\begin{figure}[!htb]
 	\centering
	\includegraphics[width=10cm]{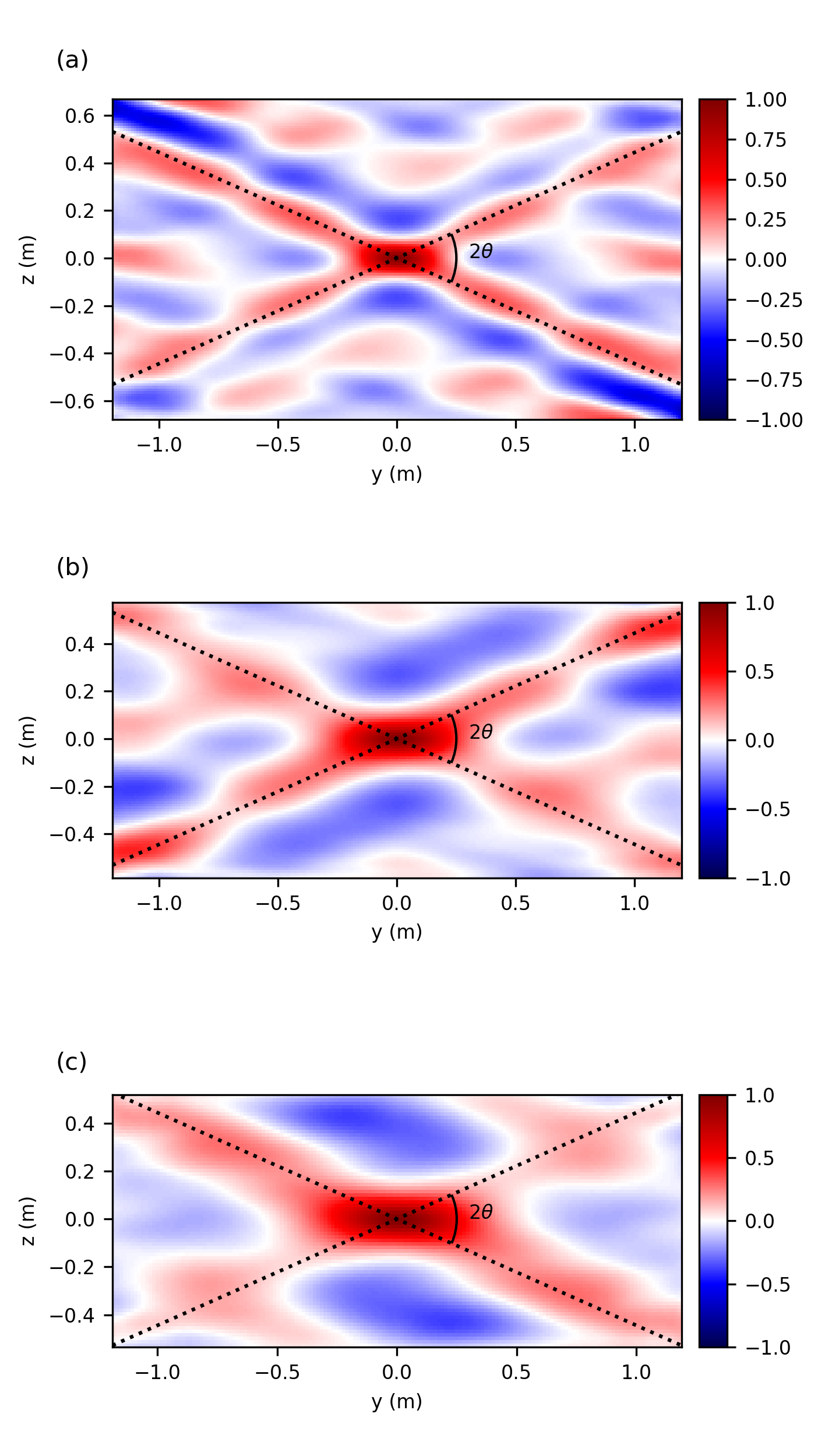} 
 	\caption{Images of vertical space-frequency correlations for horizontal component of velocity  $C^{v} (y,z,\omega)$ for various amplitudes of forcing at $\omega^*=0.406$. $\delta_i$ is the width of the central peak in the $i$ direction. (a) $A = 2\,\centi\meter$, $\delta_y = 0.14 \,\meter$, $\delta_z = 0.05\,\meter$  (b) $A = 3\, \centi\meter$, $\delta_y = 0.21\, \meter$, $\delta_z = 0.08\,\meter$ (c) $A = 4\, \centi\meter$, $\delta_y = 0.25 \,\meter$, $\delta_z = 0.09\,\meter$.}
	\label{XA}
\end{figure}

Figure \ref{XA}, shows the evolution of the St Andrew pattern at a given frequency when the forcing amplitude is increased. One can see that the width of the arms of the cross are getting wider with increasing $A$. It means that the smallest length scales present in the pattern are also increasing with $A$. This is somewhat counter-intuitive as usually the cascades proceed to smaller scales when the forcing is increased. A first interpretation may be that the small scales waves are more nonlinear and overturn and thus there is no wave structure that persists at small scales and possibly rather some sort of strongly nonlinear turbulence with eddies. This scenario would be consistent with the fact that at large forcing ($A=5$~cm) the imaging of the vertical light sheet is no longer possible due to optical index variations related to local mixing of the stratification most likely associated to overturning by small scale eddies. This interpretation is also consistent with recent numerical simulations by Yokoyama \& Nakaoka~\cite{Yokoyama:2019cm} (at similar values of the dimensionless parameters) that show that the spectral extension of the wave dominated regime is shrinking when increasing the forcing. The transition occurs at larger scales for strong forcing. A second explanation could be that small scale waves are swept by large scale motions that destroy their structure as observed for inertial waves by Campagne et al.  \cite{Campagne:2015fe} and Sharon et al. \cite{Yarom:2017ha} as well as numerically for internal waves by Minnini et al. \cite{ClarkdiLeoni:2015gh} or Yokoyama \& Nakaoka \cite{Yokoyama:2019cm}. The later authors show in their simulation that the spectral region affected by the sweeping is actually close to the border in spectral space between a region dominated by weak non linear waves and a region at smaller scales in which the dynamics is strongly nonlinear. Thus both interpretations are probably interleaved.

\begin{figure}[!htb]
 	\centering
	\includegraphics[width=16cm]{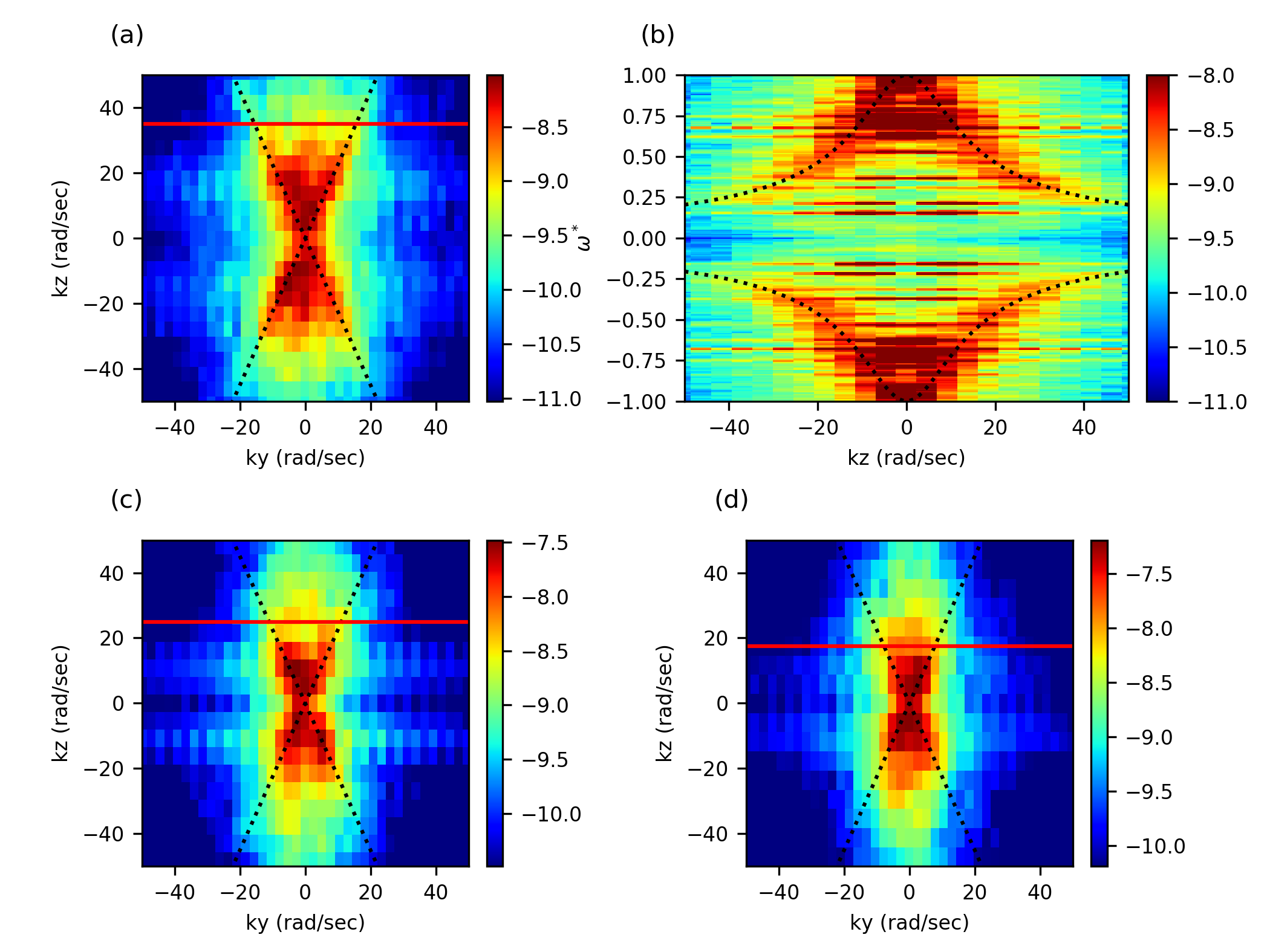}  	
	\caption{(a) Spectrum $E^v(k_y,k_z,\omega)$ for $\omega^*=0.406$ (in the continuum) and $A=2$~cm. (b) $E^{v}(k_y,k_z,\omega)+E^{w}(k_y,k_z,\omega)$ for the same experiment and $k_y=10$~rad/m. (c) \& (d) (a) Spectrum $E^v(k_y,k_z,\omega)$ at $\omega^*=0.406$ and for $A=3$~cm and $A=4$~cm respectively. The horizontal line is half the buoyancy wavenumber $k_b/2$.}
	\label{DRko}
\end{figure}

Fig.~\ref{DRko} (a) shows cuts of the spectrum for horizontal component of velocity both in frequency space and wavevector space $E^v(k_y,k_z,\omega)$ for a frequency chosen in the continuum. In principle, the information in this spectrum is the same as the one contained in the previous estimator of correlations $C^{v}$ in space of the frequency Fourier components through the Wiener-Khinchin theorem. The resolution of the Fourier transform in $k$ space is relatively poor as it is imposed by the size of the measurement domain (i.e. the size of the velocity maps in the vertical PIV which is about 0.5 meter vertically). The energy is contained in between two lines having an angle $\theta$ with the vertical as expected from the dispersion relation. The fact that energy is not localized only on the lines comes from the fact that the measurement is a 2D cut of a 3D field. Thus Fourier modes at the chosen frequency but with a $k_x$ component of the wave vector (perpendicular to the laser sheet) have a $k_y$ component which is smaller than the border imposed by the value of $\theta$. We can see that the cutoff in vertical wave number occurs at $k_z\approx 40$~rad.m$^{-1}$ which corresponds to wavelengths about 15 cm. Fig.~\ref{DRko}(b) shows a cut for $k_y = 10$~rad.m$^{-1}$. We shifted it from $k_y=0$ to avoid very large scale modes that dominate the spectrum and prevent from observing the dispersion relation. In this case, energy is spread over all frequencies lying in between the two branches of the dispersion relation.  Fig.~\ref{DRko}(c)\&(d) show the same cut as in (a) when increasing the forcing magnitude. Although the resolution in $k$ is relatively low, it can be seen that the cross pattern becomes less and less visible when increasing $A$ and that the energy is spread in a more isotropic way at large $k$. This appears consistent with the above proposed interpretation that the small scales become more nonlinear and thus the motion is no longer dominated by waves. 

The buoyancy wavenumber $k_b=N/U_h$ is associated with the vertical lengthscale of wave breaking and the vertical shear scale ($U_h$ is the order of magnitude of the horizontal velocity)~\cite{BRETHOUWER:2007bh}. As discussed above, in our case, the large scale velocity variance is not really anisotropic as in DNS, so we estimate $k_b$ as $k_b=N/u_f$. $k_b/2$ is shown in fig.~\ref{DRko} as an eye guide (horizontal red line). This value seems qualitatively consistent with the evolution of the vertical extension of energy along the dispersion relation.


\section{Concluding remarks}

In conclusion, we were able to generate a strongly stratified turbulent flow in a large scale experiment using the Coriolis facility. At the lowest forcing the Froude number is about $10^{-2}$ and the buoyancy Reynolds number close to one. In this regime we observe a wide range of frequencies and length scales for which the motion is made of weakly nonlinear waves and can most likely be called weak turbulence of internal waves. Various statistical analyses were performed to characterize the flow. It enables us to show that the spectral domain on which the weak turbulence is observed shrinks when increasing the forcing as expected from the standard phenomenology of strongly stratified turbulence. Although our flow shares similarities with this phenomenology, many features are quite different from previous observations. For instance, the frequency spectrum of horizontal velocities seem to be quite flat in contrast with oceanic observations (the Garrett \& Munk spectrum is decaying rather as $1/\omega^2$). Note that a constant frequency spectrum exists in the family of pseudo-solutions of the kinetic equation of weak internal wave turbulence by Lvov {\it et al.}~\cite{Lvov:2004p1772} although it is not observed in the ocean and although it may not be among the most likely candidates for true solutions \cite{Lvov2010}.  

The only peak that grows with $A$ at $\omega^*\approx 0.075$ is one of the gravest discrete modes with one half wavelength horizontally in the box (fig.~\ref{fig:psdamp}) and its growth may be due to an accumulation of energy due to an inverse cascade of energy (in frequency space). This accumulation is most likely stopped by viscous damping in boundary layers that prevents the formation of a strong condensate as can be observed in 2D turbulence or the inverse cascade of wave action for 2D NLS. 

The large scale anisotropy is not strongly pronounced in the experiment in contrast with numerical simulations. These discrepancies are most likely related to specificities of the experimental setup. The fact that our domain is bounded by walls is a major difference with DNS (periodic domains) and with the ocean (that can be considered as infinite in the horizontal directions). It induces the presence of large scale discrete 2D modes that most likely hides to some extent the anisotropy of the flow. The forcing scheme is also very specific to experiments. Further studies should be developed in the future to better understand these differences between experiments, DNS and the ocean. Another venue of research is to change the shape of the experimental domain to reduce the strength of the discrete strongly resonant modes.

\begin{acknowledgments}
This project has received funding from the European Research Council (ERC) under the European Union's Horizon 2020 research and innovation programme (grant agreement No 647018-WATU). We thank Bruno Voisin for his help in the computation of the correlations reported in the appendix. 
\end{acknowledgments}

\appendix\section{Computation of correlations for an axisymmetric field}
\label{appendix}
In the framework of the axisymmetric model introduced in section~\ref{F_ana}, we want to compute the spatial correlations of the components of the velocity field. 
We have (for correlations normalized by their value at the origin $\mathbf r=0$):
\begin{eqnarray}
C^u(\mathbf r,\omega)&=&\frac{\iint \langle|\mathbf a(\mathbf k,\omega)|^2\rangle\cos^2\theta\cos^2\phi \cos(\mathbf k\cdot \mathbf r) kdk\sin\theta d\phi}{\int \langle|\mathbf a(\mathbf k,\omega)|^2\rangle\cos^2\theta\cos^2\phi  kdk\sin\theta d\phi}\label{eq:A:CU}\\
C^v(\mathbf r,\omega)&=&\frac{\iint \langle|\mathbf a(\mathbf k,\omega)|^2\rangle\cos^2\theta\sin^2\phi \cos(\mathbf k\cdot \mathbf r) kdk\sin\theta d\phi}{\int \langle|\mathbf a(\mathbf k,\omega)|^2\rangle\cos^2\theta\sin^2\phi kdk\sin\theta d\phi}\label{eq:A:CV}\\
C^w(\mathbf r,\omega)&=&\frac{\iint \langle|\mathbf a(\mathbf k,\omega)|^2\rangle\sin^2\theta \cos(\mathbf k\cdot \mathbf r)  kdk\sin\theta d\phi}{\int \langle|\mathbf a(\mathbf k,\omega)|^2\rangle\sin^2\theta kdk\sin\theta d\phi}\label{eq:A:CW}
\end{eqnarray}
The integral is only on $k$ and $\phi$ as $\theta$ is imposed by the value of $\omega$ through the linear dispersion relation.

Due to axisymmetry, statistics of $\mathbf a(\mathbf k,\omega)$ are independent of $\phi$ and depend on the scalar $k$. To simplify the notations, we introduce the following notation for the weighted average of $x(k)$ over $k$ with weight $W(k)$: 
\begin{equation}
\langle x(k) \rangle_{ k}^{ W} = \frac{\int W(k) x(k) d k}{\int W(k) d k}
\end{equation}
All correlations can be written as a weighted average with $W(k,\omega)= k \langle|\mathbf a( k,\omega)|^2\rangle$:

\begin{eqnarray}
C^u(\mathbf r,\omega)&=&\left \langle \int \cos^2\phi \cos(\mathbf k\cdot \mathbf r)d\phi \right\rangle_{ k}^{W}\\
C^v(\mathbf r,\omega)&=&\left \langle \int  \sin^2\phi\cos(\mathbf k\cdot \mathbf r)d\phi \right\rangle_{ k}^{W}\\
C^w(\mathbf r,\omega)&=&\left \langle \int  \cos(\mathbf k\cdot \mathbf r)d\phi \right\rangle_{ k}^{W}
\end{eqnarray}

For the horizontal correlation, $\mathbf r=(x,y,0)$ so that $\mathbf k\cdot \mathbf r=k\sin\theta(x\cos\phi+y\sin\phi)$.  Thus

\begin{eqnarray}
C^u(x,y,\omega)=\frac{1}{\pi}\left \langle\int\cos^2\phi \cos(kx\sin\theta\cos\phi+ky\sin\theta\sin\phi) d\phi \right\rangle_{ k}^{ W}
\end{eqnarray}
and $C^v$ is the same as $C^u$ after rotation of $\pi/2$.
The integral over $\phi$ is:
\begin{eqnarray}
\int \cos^2\phi \cos(kx\sin\theta\cos\phi+ky\sin\theta\sin\phi)d\phi=2\pi\left[\frac{x^2}{x^2+y^2}J_0(\sqrt{x^2+y^2}k\sin\theta)-\frac{x^2-y^2}{x^2+y^2}\frac{J_1(\sqrt{x^2+y^2}k\sin\theta)}{   \sqrt{x^2+y^2}k\sin\theta}  \right]
\end{eqnarray}
so that

\begin{eqnarray}
C^u(x,y,\omega)=2\left \langle \left[\frac{x^2}{x^2+y^2}J_0(\sqrt{x^2+y^2}k\sin\theta)-\frac{x^2-y^2}{x^2+y^2}\frac{J_1(\sqrt{x^2+y^2}k\sin\theta)}{   \sqrt{x^2+y^2}k\sin\theta}\right]\right\rangle_{ k}^{W}
\end{eqnarray}

In the vertical plane $Oxz$, one has $\mathbf k\cdot \mathbf r=kx\sin\theta\cos\phi+kz\cos\theta$. Thus the correlation is
\begin{eqnarray}
C^u(x,z,\omega)=\frac{1}{\pi}\left \langle\int\cos^2\phi \cos(kx\sin\theta\cos\phi+kz\cos\theta) kdk d\phi\right\rangle_{ k}^{ W}
\end{eqnarray}
The integral over $\phi$ gives
\begin{equation}
\int \cos^2\phi \cos(kx\sin\theta\cos\phi+kz\cos\theta) d\phi=2\pi\left[J_0(kx\sin\theta)-\frac{J_1(kx\sin\theta)}{kx\sin\theta}\right]\cos(kz\cos\theta)
\end{equation}
so that
\begin{eqnarray}
C^u(x,z,\omega)=2\left \langle\left[J_0(kx\sin\theta)-\frac{J_1(kx\sin\theta)}{kx\sin\theta}\right]\cos(kz\cos\theta)\right\rangle_{ k}^{ W}
\label{eq:A:CUvertlong}
\end{eqnarray}

For the other components of velocity one gets similarly
\begin{eqnarray}
C^v(x,z,\omega)&=&2\left \langle\frac{J_1(kx\sin\theta)}{kx\sin\theta}\cos(kz\cos\theta) \right\rangle_{ k}^{ W}\\
C^w(x,z,\omega)&=&\left \langle J_0(kx\sin\theta)\cos(kz\cos\theta) \right\rangle_{ k}^{ W}
\end{eqnarray}

To go further in the analysis of the correlation, one must choose a model for the wave spectrum, such as $\langle\mathbf a( k,\omega)|^2\rangle\propto 1/k^\alpha$ and a range of wavenumbers to perform the $k$ integral. The lower bound of the interval is of order $k=2\pi/H$ but the highest wavenumber is relatively free (imposed in practice by the nonlinear processes and dissipation).

\begin{figure}[!htb]
 	\centering
	\includegraphics[width=\textwidth]{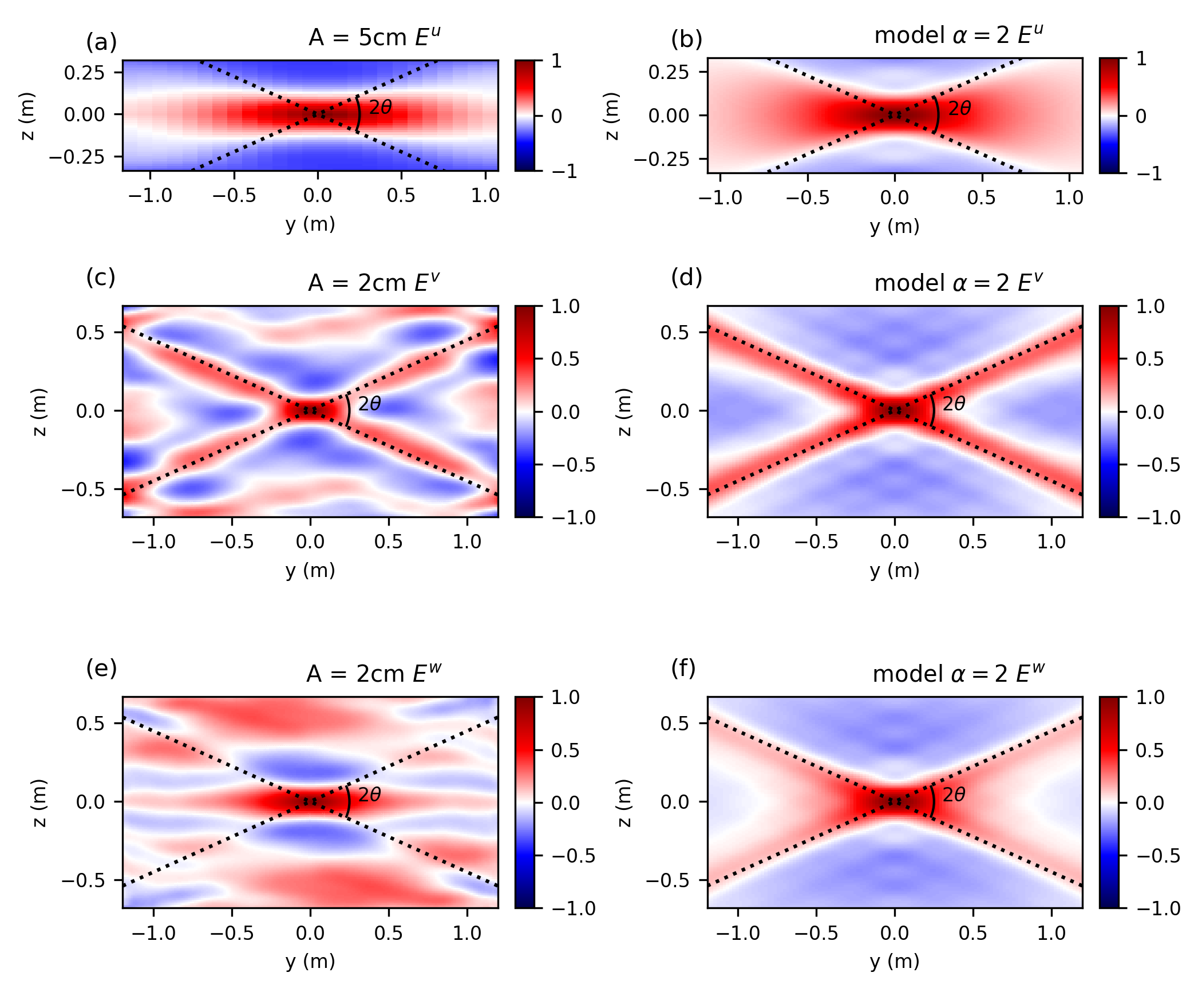}
	\caption{Example of the correlations of the 3 velocity components in the vertical plane $Oyz$ for the experiment (left column) and for the model (right) column. (a) \& (b) $C^u(y,z,\omega)$, (c) \& (d) $C^v(y,z,\omega)$, (e) \& (f) $C^w(y,z,\omega)$. The model has an interval of $k\in[4,35]$~rad.m$^{-1}$, $\alpha=2$ and $\omega^*=0.41$. (c) \& (e) come from the dataset A with PIV in the vertical laser sheet. Data in (a) comes from the dataset D with vertical scanning of the horizontal laser sheet that provides the two horizontal components of the velocity in a parallelepiped of size $3\times 2 \times 0.3$~m$^3$.}
	\label{mod0}
\end{figure}
Fig.~\ref{mod0} shows an example of correlations of the model obtained in a vertical plane $Oyz$ (for a model of the wave spectrum with $\alpha=2$) and compared to the experimental measurements. The image of $C^v(y,z,\omega)$ shows a very contrasted St. Andrew cross while that of $C^w(y,z,\omega)$ is less contrasted. The image of $C^u(y,z,\omega)$ (i.e. the component perpendicular to the chosen vertical plane) does not really show a cross but rather a horizontally elongated bump. Thus the correlation of the velocity component lying in the chosen vertical plane is the one that shows the clearest trace of the wave structure of the field. This is the one we chose to display in the analysis above for the experimental data. The experimental picture are in qualitative agreement with the model with the additional ingredient of limited statistical convergence. Its impact is most clearly visible when comparing the $C^v(y,z,\omega)$ and $C^w(y,z,\omega)$. The model shows a cross pattern on $C^w$ but much weaker than that seen in $C^v$ and in the experiment the statistical convergence is not enough to observe the cross, only the central peak is visible. It confirms that choosing the horizontal component lying in the measurement plane is the optimal choice to observe the presence of waves even with a moderate amount of data. 

\begin{figure}[!htb]
 	\centering
	\includegraphics[width=0.8\textwidth]{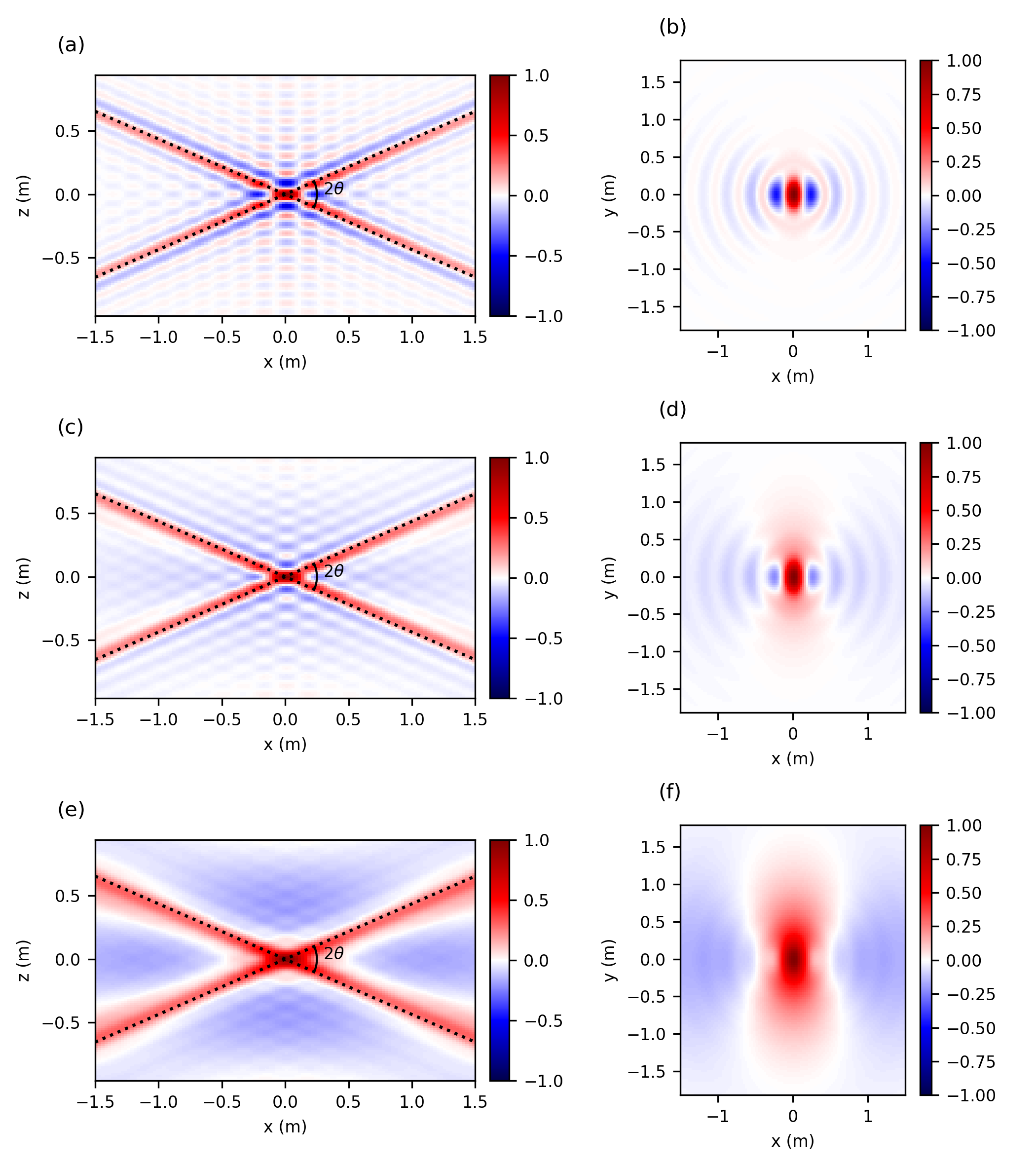}
	\caption{Impact on the choice of the wave spectrum model on the correlation $C^u (x,z,\omega)$ for $\omega^*=0.4$. (a,b) $\alpha=0$, (c,d) $\alpha=1$, (e,f) $\alpha=2$. (a,c,e) correlation in a the vertical plane $Oxz$. (b,d,f) correlation in the horizontal plance $Oxy$.}
	\label{mod1}
\end{figure}
Fig.~\ref{mod1} compares the impact of the wave spectrum decay on the vertical and horizontal correlations of the $u$ component,      for $\alpha=0,\,1,\,2$. The overall structure of the images is qualitatively unchanged. In the vertical plane, the St. Andrew cross pattern is visible in all cases and the most visible change concerns the blue parts that are slightly changing with the model. In the horizontal plane, the qualitative structure remains similar but the extension of the main positive peak of the correlation (in red) changes quite a bit with $\alpha$. Nevertheless the transverse (vertical) dimension remains larger than the longitudinal (horizontal one).

\begin{figure}[!htb]
 	\centering
	\includegraphics[width=0.9\textwidth]{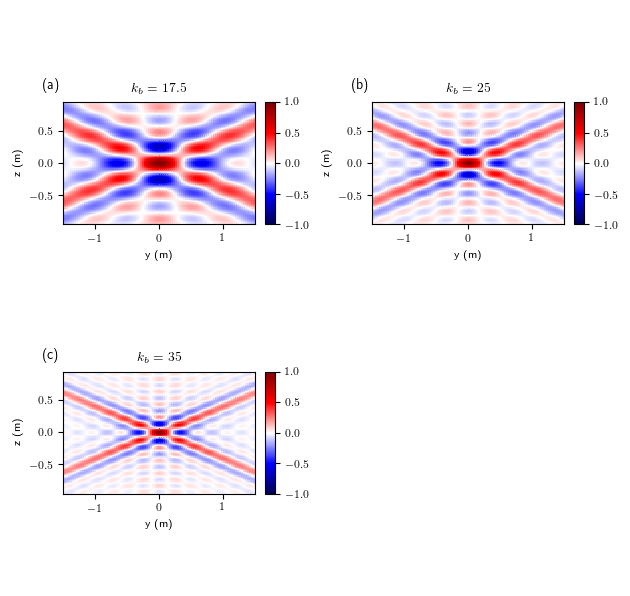} 
	\caption{Impact on the choice of the upper wavenumber $k_b$ on the vertical correlation $C^v$ for a choice of the wave spectrum with $\alpha=0$ and $\omega^*=0.4$. $C^v(y,z)$ is displayed for an easier comparison with the experimental data above and it is the same as eq. (\ref{eq:A:CUvertlong}) by symmetry. The lower bound of wavenumbers is kept constant. (a) $k_b=17.5$~rad.m$^{-1}$, (b) $k_b=25$~rad.m$^{-1}$, (c)$k_b=35$~rad.m$^{-1}$.}
	\label{mod2}
\end{figure}
Fig.~\ref{mod2} shows the impact of the choice of the upper bound of the $k$ integral in the vertical correlation. The various cases are qualitatively similar the main effect of changing the smaller scale is the width of the arms of the St. Andrew cross so it appears mostly as a scaling factor provided the scale separation between the lower and upper bounds of the $k$ interval is large enough.

\bibliography{biblio}
\end{document}